\documentclass[11pt]{article} % use larger type; default would be 10pt

\usepackage{amsmath}
\usepackage{amssymb}

\usepackage{amsthm}
\newtheorem{lemma}{Lemma}
\newtheorem{theorem}{Theorem}

 {
      \theoremstyle{plain}
      \newtheorem{assumption}{Assumption}
  }

\usepackage[utf8]{inputenc} % set input encoding (not needed with XeLaTeX)

\usepackage{siunitx}
\usepackage[left=2cm, right=5cm, top=2cm]{geometry} % to change the page dimensions
\usepackage{lscape}

\usepackage{numprint}
\usepackage{multirow}
\usepackage{subcaption}

\usepackage{mathptmx}
\usepackage{anyfontsize}
\usepackage{t1enc}

\usepackage{apacite}
\usepackage{graphicx} % support the \includegraphics command and options

\usepackage{color}
\usepackage{caption}
\usepackage{booktabs} % for much better looking tables
\usepackage{array} % for better arrays (eg matrices) in maths
\usepackage{paralist} % very flexible & customisable lists (eg. enumerate/itemize, etc.)
\usepackage{verbatim} % adds environment for commenting out blocks of text & for better verbatim
\newcommand*{\QEDA}{\hfill\ensuremath{\blacksquare}}%

\usepackage{float}
\usepackage{multirow}
\usepackage{fancyhdr}
\usepackage{datetime}

\fancypagestyle{plain}{ \fancyhf{}
 \fancyhead[L]{} \fancyhead[R]{}
 \fancyfoot[R]{Printed: \\ \today }
}

\usepackage{sectsty}
\allsectionsfont{\sffamily\mdseries\upshape} % (See the fntguide.pdf for font help)
\usepackage{setspace}
\usepackage[nottoc,notlof,notlot]{tocbibind} % Put the bibliography in the ToC
\usepackage[titles,subfigure]{tocloft} % Alter the style of the Table of Contents

\DeclareMathOperator*{\argmin}{arg\,min}

 \author{Nandana Sengupta and Fallaw Sowell} 
\begin{document}

\pagenumbering{arabic}

\title{The Ridge Path Estimator for Linear Instrumental Variables}
%\author{By Nandana Sengupta and Fallaw Sowell}
\maketitle
\thispagestyle{plain}
\begin{abstract}
\thispagestyle{plain}
This paper presents the asymptotic behavior of a linear instrumental variables (IV) estimator that uses a ridge regression penalty.  The regularization tuning parameter is selected empirically by splitting the observed data into training and test samples.  Conditional on the tuning parameter, the training sample creates a path from the IV estimator to a prior.  The optimal tuning parameter is the value along this path that minimizes the IV objective function for the test sample.  

The empirically selected regularization tuning parameter becomes an estimated parameter that jointly converges with the parameters of interest.  The asymptotic distribution of the tuning parameter is a nonstandard mixture distribution.   Monte Carlo simulations show the asymptotic distribution captures the characteristics of the sampling distributions and when this ridge estimator performs better than two-stage least squares.      
\end{abstract}
\textbf{Author Information:} 
\begin{itemize}
\begin{singlespace}
\item[1.] Nandana Sengupta (corresponding author) \\
Assistant Professor, School of Public Policy\\
Indian Institute of  Technology Delhi, India\\
\underline{email:} nandana.sengupta@sopp.iitd.ac.in \\
\underline{phone:} (+91) 9902848877 

\item[2.] Fallaw Sowell \\
Associate Professor of Economics, Tepper School of Business \\
Carnegie Mellon University, Pittsburgh, USA\\
\underline{email:} fs0v@andrew.cmu.edu \\
\underline{phone:} (+1) 412-268-3769
\end{singlespace}
\end{itemize}

\pagebreak

\setcounter{page}{1}

\section{Introduction}

% our estimator 
This paper presents the asymptotic distribution for a ridge regression estimator for the linear instrumental variable (IV) model.  The ridge estimator requires a regularization tuning parameter and can achieve lower MSE than two-stage least squares.  This estimator differs from previously studied ridge regression estimators in three important dimensions.  First, a nonzero prior.  The estimators are allowed to be shrunk towards a economically meaningful prior.  This is particularly important when the estimates are structural parameters with subject matter meaning.   Second, the regularization tuning parameter is selected empirically using the observed data.  Instead of stating asymptotic rates the tuning parameter needs to satisfy we consider a empirically selected tuning parameter and report the resulting asymptotic distribution.  

Third, the traditional GMM framework is used to characterize the asymptotic distribution of this ridge estimator.  Both adding a regularization penalty term and splitting the observed data into a training and test samples, takes the estimator out of the traditional GMM framework.  New moment conditions are presented that fit into the traditional GMM framework and include the first order conditions for the ridge estimator.

Currently, it is becoming fashionable for empirical work to use tuning parameters selected with a holdout or test sample.  However, there is a limited theoretical work on the asymptotic properties of the resulting estimators.  

The tuning parameters for ridge, Lasso and Bridge estimators are typically required to satisfy asymptotic rates of convergence to allow asymptotic results (see, \citeA{huang2008}, \citeA{Caner_2009}, and \citeA{doi:10.1080/07474938.2015.1092806}).   This leaves uncertainty because there are typically an infinite number of values that satisfy the restrictions.  In finite samples, different values for the regularization tuning parameter result in different estimates for the parameters of interest.  To avoid this indeterminacy, the observed sample is used to optimally select the value of the tuning parameter.

The ridge path estimator is the ``best'' parameter estimate over a one-dimensional path in the parameter space between the global minimum and a prior.  The global minimum is associated with low bias and high variance whereas the prior is associated with higher bias and zero variance. The trade-off between bias and variance is exploited to find the estimate with lower Mean Squared Error (MSE).  The data is split into training and test samples.  The linear IV objective function using the training sample determines the one-dimensional path and the estimate is the parameter value associated with the point on the path which minimizes the linear IV objective function using the test sample. The ridge path estimator is compared to traditional 2SLS for simulated models. We find that for low precision models with small samples, the new ridge estimator is always superior to the 2SLS estimator.  However, if the model has high precision and the sample size is large,  the ridge path estimator is competitive.

 Precision problems in linear IV estimation can occur with several models.  The past 20 years has shown a large growth in our understanding of the possible types of identification and asymptotic distributions that can occur with linear IV models (see \citeA{10.2307/23116599} for a summary): e.g. strong instruments, nearly-strong instruments, nearly-weak instrument and weak instruments. For this taxonomy, this paper and estimator is in the strong instruments setting.  A related but different model is when the number of instruments grow with the sample size (see \citeA{10.2307/2692218}).   In this paper we restrict attention to fixed number of instruments.  The models considered in this paper are closest to the situation considered in  \citeA{SANDERSON2016212}.  However, unlike \citeA{SANDERSON2016212}  we have small parameters on the instruments instead of having some of the parameters drifting to zero. In addition we focus on providing estimates for a given sample instead of testing for weak instruments.   The models we study are explicitly strongly identified, however in a finite sample the precision can be low. 
     
% background of regularization in GMM
The ridge path estimator belongs to a family of estimators which utilize regularization.  \citeA{bickel2006regularization} provides an overview of the properties of various regularization procedures in statistics. They loosely define regularization as \textit{``the class of methods needed to modify maximum
likelihood to give reasonable answers in unstable situations.''}
These estimates tend to have significantly lower variance which usually comes at the price of higher bias, i.e. the \textit{``bias-variance trade-off''}.
Nonparametric density estimation, ridge penalty estimation, LASSO penalty estimation, elastic net and spectral cutoff are all examples of regularization. For a review of methods see \citeA{hastie2009unsupervised}.

Within the structural econometrics literature, regularization concepts have recently been used by a few authors, however the intersection is still largely open. Notable contributions are the set of papers by Carrasco et al. [\citeA{carrasco2000generalization}, \citeA{carrasco2007linear}, \citeA{carrasco2012regularization}, \citeA{doi:10.1080/07474938.2015.1092806}], \citeA{caner2010adaptive} and \citeA{liao2013adaptive}. The first set of papers extend the $m$ moment conditions  to a continuum of moment conditions. The authors use ridge regularization to find the inverse of the optimal weighting \textit{operator} (instead of optimal weighting matrix in traditional GMM). \citeA{caner2010adaptive} attach a linear penalty term like in the LASSO framework  and argues that this helps by forcing parameters not significant down to zero. Finally, \citeA{liao2013adaptive} augments the $m$ moment conditions with another $k$ moment conditions where the second set of augmented moment conditions is constructed from the subset of the original $m$ moment conditions which may be misspecified. The new set of $m+k$ moment conditions and a LASSO-type penalty permit simultaneous estimation and moment selection. \citeA{doi:10.1080/07474938.2015.1092804} present a comparative analysis of different moment selection techniques via simulation studies.

The ridge path estimator extends the literature in three important dimensions. %
First, a meaningful prior is incorporated into the estimator.   
When the prior is ignored, or equivalently set to zero, the model penalizes variability about the origin.  However, in structural economic models a more appropriate penalty will be variability about some economically meaningful prior values. The parameters have meaning in the economic environment implying that prior knowledge and expertise can be incorporated by shrinking towards a prior.

Second, the data are explicitly used to select the tuning parameter.   This is in agreement with the advice to use the data in the model selection and/or tuning parameter selection. 
 Following \citeA{10.1257/jep.31.2.3} and \citeA{10.1111/ectj.12097}  we accept this sample split to determine the optimal model as a powerful tool to be embraced.  A key feature of this new estimator is splitting the sample into a training and test samples.  Lemma 1 gives the consistency and root-$n$ convergence of the empirically estimated tuning parameter.

Third, empirically selecting the tuning parameter impacts the asymptotic distribution of the parameter estimates.  As stressed in \citeA{10.2307/3533623}, the final asymptotic distribution will depend on  empirically selected tuning parameters.    We address this directly by characterizing the joint asymptotic distribution that include both the parameters of interest and the tuning parameter.  The resulting asymptotic distribution is nonstandard because the population parameter value is at the boundary of the parameter space.    We show how the ridge path estimator can be represented as a GMM estimator and are able to apply results in \citeA{andrews2002generalized}.   To our knowledge, this approach and result have not been previously presented. 

Section \ref{IV}  presents the linear IV framework, describes the precision problem and the ridge path estimator. Section \ref{asym_dist} characterizes the asymptotic distribution of the ridge path estimator in the traditional GMM framework.   Small sample properties are analyzed via simulations in Section \ref{sims}. Section \ref{conclusion} concludes. 
\section{Ridge Path Estimator for Linear Instrumental Variables Model } 
\label{IV}
This section introduces the linear IV model notation.  Ridge regression is presented as an approach to improve the MSE.  The regularization tuning parameter is empirically determined by splitting the data into training and test samples.  Conditional on the tuning parameter the ridge estimate for the training samples creates a path from the prior to the IV estimator.  The IV objective function for the test sample is then evaluated along this path to empirically determine the 
optimal tuning parameter and parameters of interest.
The asymptotic distribution of these estimates will be investigated using the GMM framework.  The first order conditions that characterize the estimates do not immediately fit into the GMM framework.  However, an alternative system of equations is presented which include the estimates.   

Consider the linear instrumental variables model where 
$Y$ is $n \times 1$,  
$X$ is $n \times k $   and 
$Z$ is $n \times m $ with $m \geq k$ 
\begin{eqnarray}
Y & = &  X \beta_0 + \varepsilon  \label{linear_model} 
\\
X & =  & Z \Gamma_{0}  + u   
\\ 
Z & = &  \left[ \begin{array}{c c c c } z_1 & z_2 &
\cdots & z_n \end{array} \right], \  z_i \sim iid, \  R_{z} = E[z_i z_i']\mbox{ is full rank,}
\\
  \mbox{ and   conditional on $Z$, } \left[ 
 \begin{array}{c}
 \varepsilon_i \\
 u_i 
 \end{array}
 \right]  & \sim & iid  \left(0, 
 \left[ 
 \begin{array}{c c} 
 \sigma^2_{\varepsilon} & \Sigma_{\varepsilon u} 
 \\
 \Sigma_{u \varepsilon} & \Sigma_U
\end{array} 
\right] \right).
\end{eqnarray}
The  IV estimator 
\begin{eqnarray}
\hat{\beta}_{IV}  & = &  \argmin_{\beta}  \frac{1}{2n} (Y- X \beta)' Z (Z' Z)^{-1} Z' (Y - X \beta) 
\label{objective}
\\
& = & 
(X' P_Z X)^{-1} X' P_Z Y \nonumber
\end{eqnarray}
where $P_Z$ is the projection matrix for  $Z$, has the asymptotic distribution
$$ 
\sqrt{n} \left( \hat{\beta}_{IV}  - \beta_0 \right) \sim_a N \left( 0, \sigma^2_\varepsilon %
\left(
\Gamma_0' R_{z} \Gamma_{0}
\right)^{-1}
\right).
$$
The covariance can be consistently estimated with 
\begin{equation}
\label{cov}
\frac{\hat{\varepsilon}' \hat{\varepsilon}}{n} \left[  \left( \frac{ X' Z}{n} \right) \left( \frac{ Z' Z}{n} \right)^{-1} 
\left( \frac{ Z' X}{n} \right) \right]^{-1}
= 
\frac{\hat{\varepsilon}' \hat{\varepsilon}}{n} \left[   \frac{ X' P_Z X}{n}  \right]^{-1}
\end{equation}
where $\hat{\varepsilon} = Y - X\hat{\beta}_{IV}$. 
Let $S_0 = E[z_{i} x_{i}' ] = R_z \Gamma_0 $.

For a finite sample let\footnote{This term is both the second derivative of the objective function (\ref{objective}) and the matrix being inverted in the last term of the covariance (\ref{cov}).}  $\frac{X' P_Z X}{n}$  have the spectral decomposition $ C \Lambda C'$, where $\Lambda$ is a positive definite diagonal $k \times k$ matrix, and $C$ is orthonormal,  $C' C = I_k$.  % Order the eigenvalues of $ \frac{X'P_{Z}X}{n}$, i.e. the diagonal elements  of $\Lambda$,  $0 < \lambda_1 \leq \lambda_2 \leq \ldots \leq \lambda_k $. 
A precision problem occurs when \textit{some} of the eigenvectors explain \textit{very little} variation, as represented by the magnitude of the corresponding eigenvalues. This occurs when the objective function is relatively flat along these dimensions and the resulting covariance estimates are large because as equation (\ref{cov}) shows, the variance of $\hat{\beta}_{IV}$ is proportional to  $  \left( \frac{X' P_Z X}{n} \right)^{-1} = \left( C \Lambda C'\right)^{-1}  = C \Lambda^{-1} C'.$
The flat objective function, or equivalently large estimated variances, leads to a relatively large MSE.  The ridge path estimator addresses this problem by shrinking the estimated parameter toward a prior.   The IV estimate still has \textit{low bias} (it is consistent) and has the \textit{asymptotically minimum variance}.  However, accepting a little higher bias can have a dramatic reduction in the variance and thus provide a point estimate with lower MSE.  

%\subsection{The path associated with Ridge Regression for Linear IV}

The ridge objective function augments the usual  IV objective function (\ref{objective}) with a quadratic penalty centered at a prior value, $\beta^p$, weighted by a regularization tuning parameter $\alpha$ 
\begin{eqnarray}
J_n  (\beta) & = &\frac{1}{2n} (Y- X \beta)' P_Z (Y - X \beta)  + \frac{1}{2} \alpha (\beta - \beta^p )' (\beta - \beta^p). 
\label{ridge_objective}
\end{eqnarray}
The objective function's second derivative is
$ \left(\frac{X'  P_Z  X}{n} + \alpha I_k \right)  = C(\Lambda + \alpha I_k )C'.
$
The regularization parameter injects stability since
$
 \left( \frac{X' P_Z X}{n} + \alpha I_k\right)^{-1}  =  C  \left(\Lambda + \alpha I_k \right) ^{-1} C' 
$
has eigenvalues  $ 1/(\lambda_i + \alpha) $ for $i=1, \ldots, k$ which are decreasing in $\alpha$.  This results in smaller variance but higher bias. 

Denote the ridge solution given $\alpha$ as
\begin{eqnarray}
\hat{\beta}_{IV} (\alpha) & = & \left( \frac{X' P_Z X}{n}  + \alpha I_k\right)^{-1}\left( \frac{X' P_Z Y}{n}
+  \alpha \beta^p \right) \label{evaluation} \nonumber \\
           & = &C \left( \Lambda + \alpha I_{k}\right)^{-1} C' \frac{X'P_ZY}{n}
+  C \left( \Lambda + \alpha I_{k}\right)^{-1} C'\alpha \beta^p  \nonumber \\
  & = & C \left( \Lambda + \alpha I_{k}\right)^{-1} C' \cdot \left[ C\Lambda C' \cdot C \Lambda^{-1} C' \right]  \frac{X'P_ZY}{n}
  + C \left( \frac{\Lambda}{\alpha} + I_{k}\right)^{-1} C' \beta^p \nonumber  \\
%                                   & =&C \left( \Lambda + \alpha I_{k}\right)^{-1} C' \cdot C\Lambda C'   \left( \frac{X'X}{n} \right)  \frac{X'Y}{n}
%+ C \left( \frac{\Lambda}{\alpha} + I_{k}\right)^{-1} C' \beta^p \\
%
  & =&  C \left( I_{k} + \alpha \Lambda^{-1}\right)^{-1} C'  \hat{\beta}_{IV} + C \left( \frac{\Lambda}{\alpha} + I_{k} \right)^{-1} C' \beta^p. \label{path}
\end{eqnarray}
Equation (\ref{path}) shows how the tuning parameter, $\alpha$ creates a  smooth curve in the parameter space between the \textit{low bias-high variance} IV estimate, $\hat{\beta}_{IV}$, (when $\alpha = 0$) to the \textit{high bias-no variance} prior, $\beta^p$, (when $\alpha \rightarrow \infty$).  The ridge estimator should be evaluated using equation (\ref{evaluation}) because the IV estimator is poorly defined for the situations considered in this paper. 

%\subsection{Implementing the Ridge Path Estimator (Training and Test Samples in Linear IV)}
%
Different values of $\alpha$ result in different values of $\beta$. The optimal value of $\alpha$ is determined empirically as follows.
The data are split into training and test samples. The training sample is the first $[\tau n]$ observations, denoted, $Y_{\tau n}$, $X_{\tau n}$, and $Z_{\tau n}$, and are used to calculate a path between the IV estimate and the prior as in equation (\ref{evaluation}). The estimate using the training sample, conditional on $\alpha,$ is 
\begin{eqnarray}
\hat{\beta}_{IV, \tau n}(\alpha) & \equiv & \argmin_{\beta}  \frac{1}{2 [\tau n]}  \left( Y_{\tau n} - X_{\tau n} \beta \right)'
P_{Z_{\tau n}} 
\left( Y_{\tau n} - X_{\tau n} \beta \right) + \frac{\alpha}{2}  (\beta - \beta^p)' (\beta - \beta^p) 
\end{eqnarray}
where $P_{Z_{\tau n}}  $ is the projection matrix onto 
$Z_{\tau n}$ and  $[\cdot]$ is the greatest integer function. 
The first order conditions for an internal solution are
\begin{eqnarray}
 -\frac{1}{ \tau n} X_{\tau n}'
P_{Z_{\tau n}}
\left( Y_{\tau n} - X_{\tau n} \hat{\beta} \right) + \alpha ( \hat{\beta} - \beta^p) & = & 0  \nonumber
\end{eqnarray}
or alternatively
\begin{eqnarray}
 -\frac{1}{ [\tau n]} \sum_{i=1}^{[\tau n]}  \left\{ \left(  \frac{X_{\tau n}'
Z_{\tau n} }{  [\tau n] } \right) \left( \frac{Z_{\tau n}' Z_{\tau n} }{ [\tau n]} \right)^{-1} \right\} z_i
\left( y_i - x_i' \hat{\beta} \right) 
+ %\Bigg\{ 
\alpha ( \hat{\beta} - \beta^p) %\Bigg\} 
& = & 0. 
\label{beta_FOC}
\end{eqnarray}
The closed form solution is 
\begin{eqnarray}
\hat{\beta}_{IV, \tau n}(\alpha) 
& = & \left( 
\frac{X_{\tau n}'
P_{Z_{\tau n}} 
 X_{\tau n}}{[\tau n]} 
+ \alpha I \right)^{-1} \left( \frac{X_{\tau n}'
P_{Z_{\tau n}} 
 Y_{\tau n} }{[\tau n]}
  + \alpha \beta^p \right).  
\end{eqnarray}
As $\alpha$ goes from 0 towards  infinity, this gives a path from the IV estimator, $\hat{\beta}_{IV, \tau n}$ (at $\alpha = 0$), to the prior, $\beta^p$ (the limit as $\alpha \rightarrow \infty$).
Following this path, the optimal $\alpha$ is selected to minimize the IV least squares objective function (\ref{objective}) over the remaining $(n - [\tau n])$ observations, the test sample, denoted $Y_{n(1-\tau)}$, $X_{n(1-\tau)}$ and $Z_{n(1-\tau)}$.   The optimal value for the tuning parameter is defined by $ \hat{\alpha} = \argmin_{\alpha \in [0, \infty)}  Q_{n(1- \tau)}(\alpha)$
where
\begin{equation}
\label{alpha_equation}
Q_{n(1-\tau)} (\alpha) = 
 \frac{1}{2(n- [n\tau])} \left(Y_{n(1-\tau)} - X_{n(1-\tau)} \hat{\beta}_{IV, \tau n}(\alpha) \right)' 
P_{Z_{n(1-\tau)}}
\left(Y_{n(1-\tau)} - X_{n(1-\tau)} \hat{\beta}_{IV, \tau n}(\alpha) \right)
\end{equation}
where $P_{Z_{n(1-\tau)}}$ is the projection matrix onto  $Z_{n(1-\tau)}.$ 
The first order condition for an internal solution is 
\begin{eqnarray}
 \frac{1}{(n - [\tau n])} (\beta^p -\hat{\beta}_{IV,\tau n}(\hat{\alpha}) )' \left( \frac{X_{\tau n}' P_{Z_{\tau n}} X_{\tau n}}{ [\tau n] }  + \hat{\alpha} I_k\right)^{-1}  
 X_{n(1-\tau)}' 
P_{Z_{n(1-\tau)}}
\left(Y_{n(1-\tau)} - X_{n(1-\tau)} \hat{\beta}_{IV, \tau n}(\hat{\alpha}) \right)
& = & 0 \nonumber
\end{eqnarray}
or alternatively
\begin{eqnarray}
 \frac{1}{n - [\tau n]} \sum_{i= [\tau n] + 1}^{n}  \Bigg\{ (\beta^p -\hat{\beta}_{IV,\tau n}(\hat{\alpha}) )' \left( \left(  \frac{X_{\tau n}'
Z_{\tau n} }{  [\tau n] } \right) \left( \frac{Z_{\tau n}' Z_{\tau n} }{ [\tau n]} \right)^{-1} \left(  \frac{X_{\tau n}'
Z_{\tau n} }{  [\tau n] } \right) + \hat{\alpha} I_k\right)^{-1}  
\nonumber
\\
 \left(  \frac{X_{\tau n}'
Z_{\tau n} }{  n - [\tau n] } \right) \left( \frac{Z_{\tau n}' Z_{\tau n} }{ n - [\tau n]} \right)^{-1} 
\Bigg\}
z_i \left(y_i - x_i' \hat{\beta}_{IV, \tau n}(\hat{\alpha}) \right)
& = & 0.
\label{alpha_FOC}
\end{eqnarray}
The ridge path regression estimate is $\hat{\beta}_{ \hat{\alpha}} \equiv \hat{\beta}_{IV, \tau n}( \hat{\alpha})$. 

The first order conditions that characterize the ridge path estimator, equations (\ref{beta_FOC}) and (\ref{alpha_FOC}), are $k+1$ equations in the $k+1$ parameters and have the structure of sample averages being set to zero. However, the functions being averaged do not fit into the traditional GMM framework.  In equations (\ref{beta_FOC}) and (\ref{alpha_FOC}) the terms in the curly brackets depend on the entire sample and not just the data for index $i$ and the parameters.  The terms in the curly brackets will converge at ${O}_p \left( n^{-1/2} \right)$ and must be considered jointly with the asymptotic distributions of 
$ (\hat{\beta}_{IV, \tau n}(\hat{\alpha})', \hat{\alpha})'$. 

The asymptotic distribution of the ridge path estimator can be determined with the GMM framework using the parameterization 
 $\theta = \left[ 
\begin{array}{c c c c c c}
{\rm vech}(R_\tau)' & {\rm vec}(S_\tau)'   & \beta' & \alpha &   {\rm vech} \left( R_{(1-\tau)} \right)' & {\rm vec} \left(S_{(1-\tau)} \right)' 
\end{array}
\right]'$ 
where 
${\rm vec}( \cdot)$ stacks the elements from a matrix into a column vector
and 
${ \rm vech}( \cdot)$ stacks the unique elements from a symmetric matrix into a column vector.  The population parameter values are
$$
\theta_0 = \left[ 
\begin{array}{c c c c c c}
{\rm vech}(R_z)' & {\rm vec}( R_z \Gamma_{0} )' & \beta_{0}'  & 0 & {\rm vech}(R_z)' & {\rm vec}( R_z \Gamma_0 )' \end{array}
\right]'.
$$
The ridge path estimator is part of the parameter estimates defined by the just identified system of equations 
\begin{equation}
\label{moment_conditions}
H_n(\theta) = \frac{1}{n} \sum_{i=1}^{n}  \left[ 
\begin{array}{c}
{\bf 1}_{\tau n}(i) {\rm vech}(R_\tau - z_i z_i') \\
{\bf 1}_{\tau n}(i) {\rm vec}(S_\tau - z_i x_i') \\
{\bf 1}_{\tau n}(i)  \left( S_\tau' R_\tau^{-1} z_i ( y_i  - x_i' \beta)  + \alpha (\beta - \beta^p ) \right) \\
(1 - {\bf 1}_{\tau n}(i) ) (y_i - x_i '\beta) z_i 'R_{(1-\tau)}^{-1} S_{(1-\tau)} \left( S_\tau' R_\tau^{-1} S_\tau + \alpha I_k \right)^{-1} (\beta^p - \beta)
\\
(1 - {\bf 1}_{\tau n}(i) ) {\rm vech}(R_{(1-\tau)} - z_i z_i') \\
(1 - {\bf 1}_{\tau n}(i) )  {\rm vec}(S_{(1-\tau)} - z_i x_i') 
\end{array}
\right]
\end{equation}
where the training and test samples are determined with the indicator function 
$$
 {\bf 1}_{\tau n}(i) = \left\{ \begin{array}{l l} 1, & i \leq [\tau n] 
 \\
 0, &  [\tau n] < i.  \end{array} \right.
$$

\section{Asymptotic Behavior} 
\label{asym_dist}

Three assumptions are sufficient to obtain asymptotic distribution for the ridge path estimator.
\begin{assumption}
$z_i$ is iid  with finite fourth moments and $E[z_i z_i'] = R_z$ has full rank. 
\end{assumption}

\begin{assumption}
Conditional on $Z$,
$ \left[ 
 \begin{array}{c c}
 \varepsilon_i 
 &
 u_i' 
 \end{array}
 \right]'$ are iid vectors with zero mean,  full rank covariance matrix with possibly nonzero off-diagonal elements. 
\end{assumption}

{\bf  Assumptions 1} and {\bf 2} imply $E[h_i(\theta_0)] = 0$ and
$\sqrt{n} H_n(\theta_0)$ satisfies the CLT.

\begin{assumption} The parameter space $\Theta$ is defined by: $\
R_z$ is restricted to a symmetric positive definite matrix with eigenvalues $  1/B_1 \leq \tilde{\lambda}_1 \leq \tilde{\lambda}_2 \leq  \ldots \leq \tilde{\lambda}_m \leq B_1,$ 
$ \left| \beta_j \right| \leq B_{2} $ for $j = 1, 2, \ldots, k$, $\Gamma_0 = [ \gamma_{\ell,j}  ]$ is of full rank with 
$ \left| \gamma_{\ell,j} \right| \leq B_{3} $ for $\ell = 1, \ldots, m$, $j = 1, 2, \ldots, k$ 
and 
$\alpha \in [0, B_4]$ where $B_1$, $B_2$, $B_{3}$ and $B_4$ are positive and finite.

\end{assumption}

First consider the tuning parameter.  Even though it is empirically selected using the training and samples, its limiting value and rate of convergence are familiar. 

\begin{lemma} Assumptions 1, 2 and 3 imply 
\begin{enumerate}
\item $\hat{\alpha} \rightarrow 0$  and 
\item $ \sqrt{n} \hat{\alpha} = O_p(1). $

\end{enumerate}

\end{lemma}  
\noindent

\noindent
Proofs are given in the appendix. 

Lemma 1 implies that the population parameter value for the tuning parameter is zero, 
$ \alpha_0 = 0$, which is on the boundary of the parameter space.  This results in a nonstandard asymptotic distribution which can be characterized by appealing to {\bf Theorem 1} in \citeA{andrews2002generalized}.   
The approach in \citeA{andrews2002generalized} requires the root-$n$ convergence of the parameters.  Lemma 1, traditional 2SLS and method of moments establishes this for all the parameter in $\theta$. 
Equation (\ref{moment_conditions}) puts the ridge path estimator in the form of the first part of equation (14) from  \citeA{andrews2002generalized}.
Because the system is just identified, the weighting matrix does not affect the estimator and is set to the identity matrix. 
The scaled GMM objective function can be expanded into a quadratic approximation about the centered and scaled population parameter values
\begin{eqnarray*}
 n H_n(\theta)' H_n(\theta) 
& = & nH_{n}(\theta_0)' H_n(\theta_0)   + nH_n(\theta_0) \frac{\partial H_n(\theta_0)}{\partial \theta'} (\theta - \theta_0)  \\
& & \hspace{.3in}
 + \frac{n}{2} (\theta - \theta_0)' \left\{ 
\frac{\partial H_n(\theta_0)'}{\partial \theta} 
\frac{\partial H_n(\theta_0)}{\partial \theta'} 
 \right\} (\theta - \theta_0) + o_p( 1 ) 
 \\
& = & \frac{n}{2} H_{n}(\theta_0)' H_n(\theta_0) \nonumber
%\\
%&  &
 +  
\frac{n}{2} \left( H_n(\theta_0) + \frac{\partial H_n(\theta_0)}{\partial \theta'} (\theta - \theta_0) \right)'  
 \left( H_n(\theta_0) + \frac{\partial H_n(\theta_0)}{\partial \theta'} (\theta - \theta_0) \right)
\nonumber 
 + o_p( 1 ) 
 \end{eqnarray*}
 \begin{eqnarray*}
& = & \frac{n}{2} H_{n}(\theta_0)' H_n(\theta_0) 
\nonumber
%\\
%&  & 
+  
\frac{1}{2} \left( \left(-  \frac{\partial H_n(\theta_0)}{\partial \theta'}  \right)^{-1} \sqrt{n} H_n(\theta_0) - \sqrt{n}(\theta - \theta_0) \right)'  
 \left\{ 
\frac{\partial H_n(\theta_0)'}{\partial \theta} 
\frac{\partial H_n(\theta_0)}{\partial \theta'} 
 \right\} 
 \nonumber
 \\
& & \hspace{.5in} \times \left( \left(-  \frac{\partial H_n(\theta_0)}{\partial \theta'}  \right)^{-1}  \sqrt{n}H_n(\theta_0) - \sqrt{n} (\theta - \theta_0) \right)
\nonumber + o_p( 1 ).  
\end{eqnarray*}

The first term does not depend on $\theta$ and the last term converges to zero in probability.  This suggests  selecting $\hat{\theta}$ to minimize  $H_n(\theta)' H_n(\theta)$ will result in the asymptotic distribution of $  \sqrt{n}( \hat{\theta} - \theta_0) $ being the same as the distribution of  $\lambda \in 
\Lambda \equiv \left\{ \lambda \in R^{ m(m+1) + 2 km  + k+1}: \lambda_{ m(m+1)/2 + km + k + 1} \geq 0 \right\}
$ where $ ( {\cal Z} - \lambda)' M_0' M_0 ({\cal Z} - \lambda) $ takes its minimum, where the random variable is defined as 
\begin{eqnarray*}
% & & Z
% \\ 
& & {\cal Z}   =  \lim_{n \rightarrow \infty} \left( E \left[ -  \frac{\partial H_n(\theta_0)}{\partial \theta'} \right] \right)^{-1} \sqrt{n} H_n(\theta_0)
\end{eqnarray*}
and
$$
M_0 = E \left[   \frac{\partial H_n(\theta_0)}{\partial \theta'} \right] .
$$
This indeed is the result by {\bf Theorem 1} of \citeA{andrews2002generalized}.
The needed assumptions are given in \citeA{andrews2002generalized}.     
The estimator is defined as 
$$
\hat{\theta} =  
\argmin_{\theta \in \Theta }  \hspace{.1in} H_n(\theta)' H_n(\theta).  
$$

\begin{theorem}   {\bf Assumptions} {\bf 1 - 3} imply the asymptotic distribution of  $  \sqrt{n}( \hat{\theta} - \theta_0) $  is equivalent to the distribution of
$$
\hat{\lambda} = \argmin_{\lambda \in \Lambda} \hspace{.2in} ( {\cal Z} - \lambda)' M_0' M_0 ({\cal Z} - \lambda).
$$
\end{theorem}

%\subsection{Intuition}

The objective function can be minimized at a value of the tuning parameter in $(0, \infty)$ or possibly at $\alpha = 0.$      The asymptotic distribution of the tuning parameter will be composed of two parts, a discrete mass at $\alpha = 0 $ and a continuous function over $(0,\infty)$.
The asymptotic distribution over the other
parameters can be thought of as being composed of two parts, the distribution conditional on $\alpha = 0$ and the distribution over $ \alpha > 0.$ 

In terms of the framework presented in \citeA{andrews2002generalized}, the random sample is used to create a random variable.  This is then projected onto the parameter space, which is a cone.    The projection onto the cone results in the discrete mass at $\alpha = 0$ and the continuous mass over $(0, \infty)$.  
As noted in \citeA{andrews2002generalized}, this type of a characterization of the asymptotic distribution can be easily programmed and simulated.

\section{Small Sample Properties} \label{sims}
   
To investigate the small sample performance, linear IV models are simulated and estimated using 2SLS and the ridge path estimator.  The model is given in equations (1) to (4) with $k=2$ and $m=3$.   
To standardize the model,  set $z_i \sim \mbox{iid} N(0, I_3)$ and $\beta_{0}$ = (0, 0)'. Endogeneity is created with  
$$ 
\left[ \begin{array}{c } \varepsilon_i \\  u_i \end{array} \right] \sim 
\mbox{iid} N\left( 0, \left[ \begin{array}{c c c} 1 & .7 & .7 \\
                                                  .7 & 1 & 0 \\
                                                  .7 & 0 & 1 
                                                  \end{array} \right] \right). 
$$
The strength of the instrument signal is controlled by the parameter\footnote{Similar results are obtained via other specifications of $\Gamma_0$. These are included as part of supplementary material for the paper, available from the authors on request.} $\delta$ in  
$$
\Gamma_0 = \begin{bmatrix}
            1  & 0 \\
            0 & \delta \\
            1 & 0 
            \end{bmatrix}.
$$
To judge the behavior of the estimator, three different dimensions of the model are adjusted. 

\begin{enumerate}

\item  Sample size. For smaller sample sizes, the ridge path estimator should have better properties whereas for larger sample sizes, 2SLS should perform better. We consider sample sizes of $n = 25$, 50, 250 and 500. 

\item Precision. Signal strength of the instruments is one way to vary precision. The instrument signal strength decreases with the value of $\delta$ above, conditional on holding the other model parameters fixed. For lower precision settings or smaller signal strengths the ridge path estimator should perform better. We consider values of $\delta = 0.1$, 0.25, 0.5 and 1. Note that while $\delta = 1$ leads to a high precision setting for all sample sizes considered, $\delta = 0.1$ leads to a low precision setting in smaller samples and a high precision setting in larger samples.  

\item Prior value relative to $\beta_0$. For the prior closer to the population parameter values the ridge path estimator should perform relatively better. We consider values of $\beta^p$ which were a) one standard deviation\footnote{Each individual error term is standard normal.}  from the true value $\beta^p = (1/\sqrt{2}, 1/\sqrt{2})'$, b) two standard deviations from the true value $\beta^p = (\sqrt{2}, \sqrt{2})',$  and c) three standard deviations from the true value\footnote{Other specifications of prior values also led to similar results. These are included as part of supplementary material for the paper, available from the authors on request.} $\beta^p = (3/\sqrt{2}, 3/\sqrt{2})'$. 

\end{enumerate}

%The simulated models are presented to demonstrate the basic features that determine the relative performance of the estimators: sample size, precision, and priors. Because the models are all identified, as the sample size increase the 2SLS estimator should do better with no need for any regularization. For a fixed samples size the lower the precision of the model, the better the ridge path estimators are expected to perform. Finally, the relative performance of the estimators also depends on the selection of the priors. A prior equal to the population parameter value gives an unfair advantage to the ridge path estimators whereas a poorly specified prior can bias the ridge path estimator.  

We simulate a total of $48$ model specifications corresponding to $4$ sample sizes $n$, $4$ values of the precision parameter $\delta$  and $3$ values of the prior $\beta^p$. 
Each specification is simulated $10,000$ times and both 2SLS and ridge path estimator are estimated. We compare estimated $\beta_0$ values on bias, variance and MSE. For the ridge path estimator we use $\tau = .7$ to split the sample between training and test samples. 

The regularization parameter $\alpha$ is selected in two steps -- first, we search in the log-space going from $10^{-5}$ to $10^6$; second, we perform a grid search\footnote{We consider a linear grid of $10,000$ points in the the second step.} in a linear space around the value selected in the first step. A final selected value of $\hat{\alpha} = 0$ in the second step corresponds to a ``no regularization" scenario which implies the ridge path estimator ignores the prior in favor of the data and the value $\hat{\alpha} = 10^7$ corresponds to an ``infinite regularization" scenario which implies the ridge path estimator ignores the data in favor of the prior. 

Tables \ref{tabBeta1} and \ref{tabBeta3} compare the performance of the 2SLS estimator with the ridge path estimator for different precision levels and sample sizes when the prior is fixed at $\beta^p = (\frac{1}{\sqrt{2}}, \frac{1}{\sqrt{2}})'$ and $\beta^p = (\frac{3}{\sqrt{2}}, \frac{3}{\sqrt{2}})'$ respectively. Recall, our parameter of interest is $\beta_0 = (\beta_1, \beta_2)' =  (0,0)'$. We compare the estimators based on a) bias, b) standard deviation of the estimates, c) MSE values of the estimates and d) sum of MSE values of $\hat{\beta_1}$ and $\hat{\beta_2}$. In both tables, the 2SLS estimator performs as expected --  \textit{both} bias and standard deviation of estimates fall as sample size increases and as instrument signal strength increases. In smaller samples, the 2SLS estimators exhibit some bias, which confirms that 2SLSL estimators are consistent but not unbiased. Table \ref{tabBeta1} presents a scenario where the prioir for the ridge path estimator is one standard deviation away from the true parameter estimate. We note that in the low precision setting of $\delta= 0.1$ the ridge path estimator has lower MSE for all sample sizes considered in the simulations. However as precision improves, we note that for larger sample sizes the 2SLS estimator has lower MSE. Table \ref{tabBeta3} describes a scenario where the ridge path estimator does not have any particular advantage since it is biased to a prior which is $3$ standard deviations away from the true parameter value. However, even when prior values are far from true parameter values, there are a number of scenarios where the ridge path estimator outperforms the 2SLS estimator in terms of  MSE. In particular, in small samples and low precision settings, the ridge path estimator leads to smaller MSE. When $\delta = 0.1$, the ridge path estimator leads to lower MSE values for all sample sizes except $n = 500$. When  $\delta = 1$ and the model has high precision, the ridge path estimator has higher MSE than 2SLS. Thus as the signal strength improves and low precision issues subside, 2SLS dominates. The bias-variance trade-off is at work here. Consider the results corresponding to $n = 25$ and $\delta = 0.25$. The ridge path estimator has \textit{higher} bias compared to the 2SLS estimator for both parameters, however this is compensated by considerably smaller standard deviation values leading to smaller MSE. This table also demonstrates scenarios where for a given $\delta$ value, as the sample size increases the estimator with lower MSE changes from ridge path to 2SLS. For $\delta = 0.25$, the ridge path estimator performs better for sample sizes $n\leq 50$ whereas 2SLS performs better for $n\geq 250$. Similarly, for $\delta = 0.50$, the ridge path estimator outperforms 2SLS only for the smallest sample size of $n =25$. 

\begin{figure}[H]
\caption{ Scatter plots of the estimates from 2SLS and ridge path estimator with different priors when precision is low ($\delta = 0.1$) and sample size is small ($n = 25$). Estimates, the true parameter value and prior values are represented by blue, yellow and red points respectively.  The variance associated with 2SLS estimates is much larger than the ridge path estimates. The ridge path estimator is biased toward the prior. }\label{dlt10n25}

%On the other hand, for high precision models ($\delta = 1$)the variance reduction from 2SLS for the ridge path estimator is not quite so dramatic. In fact while the variance reduction appears substantial for prior values $\beta^p = (0,0)$ and $\beta^p = (\frac{1}{\sqrt{2}} , \frac{1}{\sqrt{2}})$, it is not clear at least visually if there is a reduction in variance for a poorly specified prior at $\beta^p = (\frac{3}{\sqrt{2}} , \frac{3}{\sqrt{2}})$. In larger samples we note that except when $\beta^p = (0,0)$, the 2SLS estimates outperform the ridge path estimators which is demonstrated by larger clouds which are slightly off-center from the true parameter values. However, we also note that the ridge path estimators using different priors are still competitive and don't lead to a drastically worse performance
\centering
\includegraphics[width=0.95\textwidth]{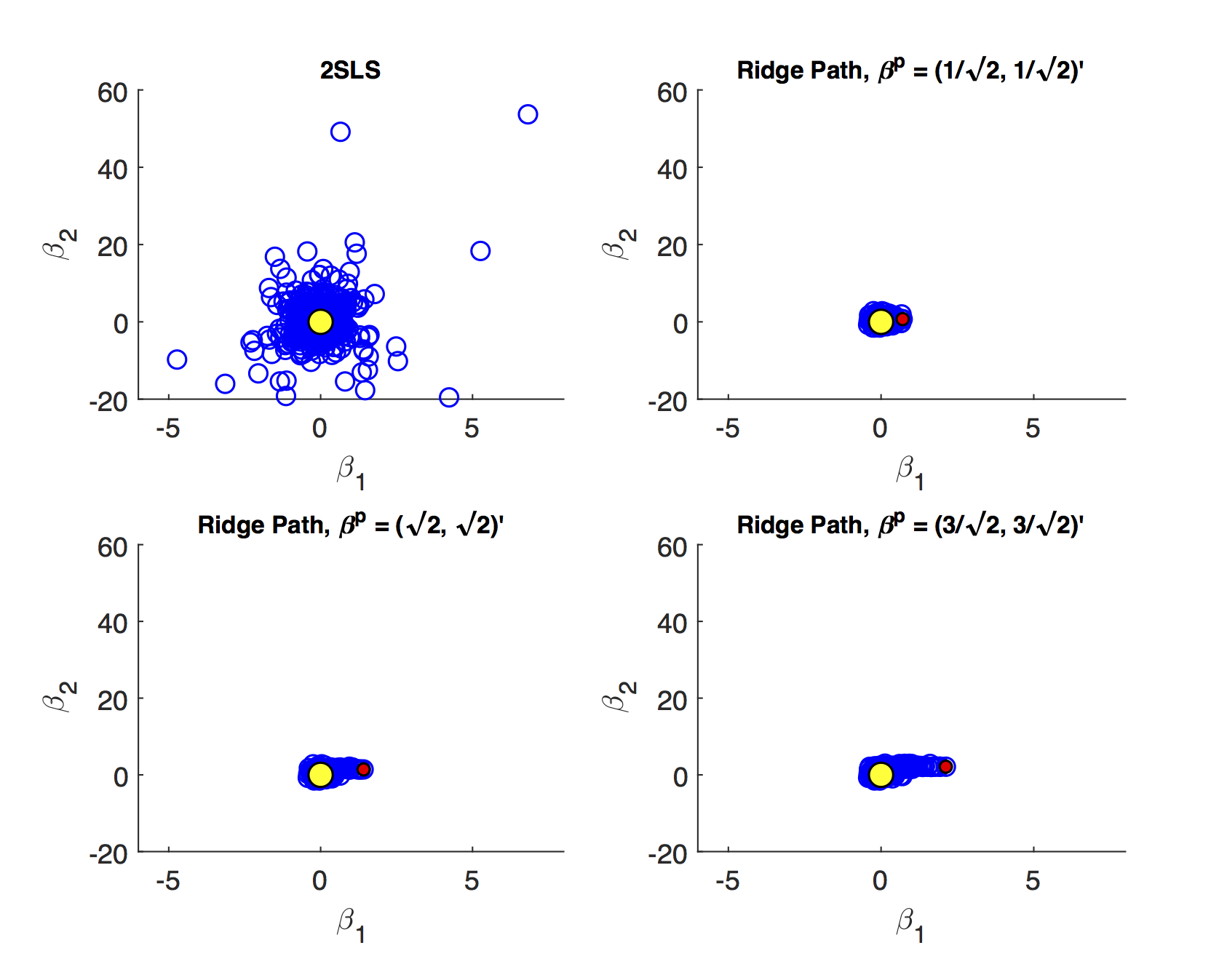}
\end{figure}

\begin{figure}[H]
\caption{Scatter plots of the estimates from 2SLS and ridge path estimator with different priors when precision is low ($\delta = 0.1$) and sample size is large ($n = 500$). Estimates, the true parameter value and prior values are represented by blue, yellow and red points respectively.  The variance associated with 2SLS estimates is much larger than the ridge path estimates. The ridge path estimator is less biased towards the prior in the larger samples, but we note that especially in the case where $\beta^p = (\frac{3}{\sqrt{2}}, \frac{3}{\sqrt{2}})'$, estimates are being pulled toward the prior.}\label{dlt10n500}
\centering
\includegraphics[width=0.95\textwidth]{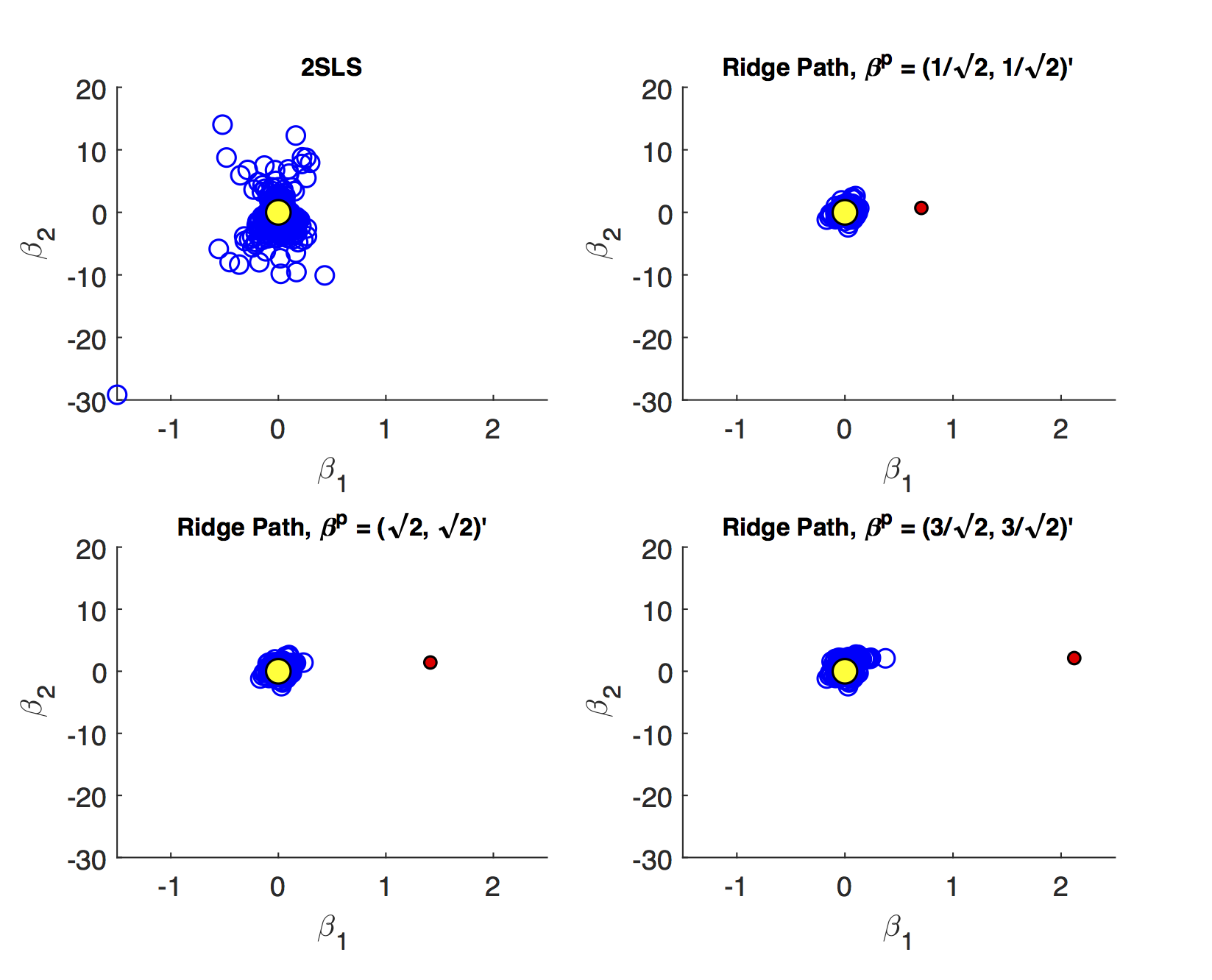}
\end{figure}

\begin{figure}[H]
\centering
\caption{Scatter plots of the estimates from 2SLS and ridge path estimator with different priors when precision is high ($\delta = 1$) and sample size is small ($n = 25$). Estimates, the true parameter value and prior values are represented by blue, yellow and red points respectively. 2SLS performance is much better in this setting. The variance reduction for the ridge path estimator is not as dramatic. }\label{dlt100n25}
\includegraphics[width=0.95\linewidth]{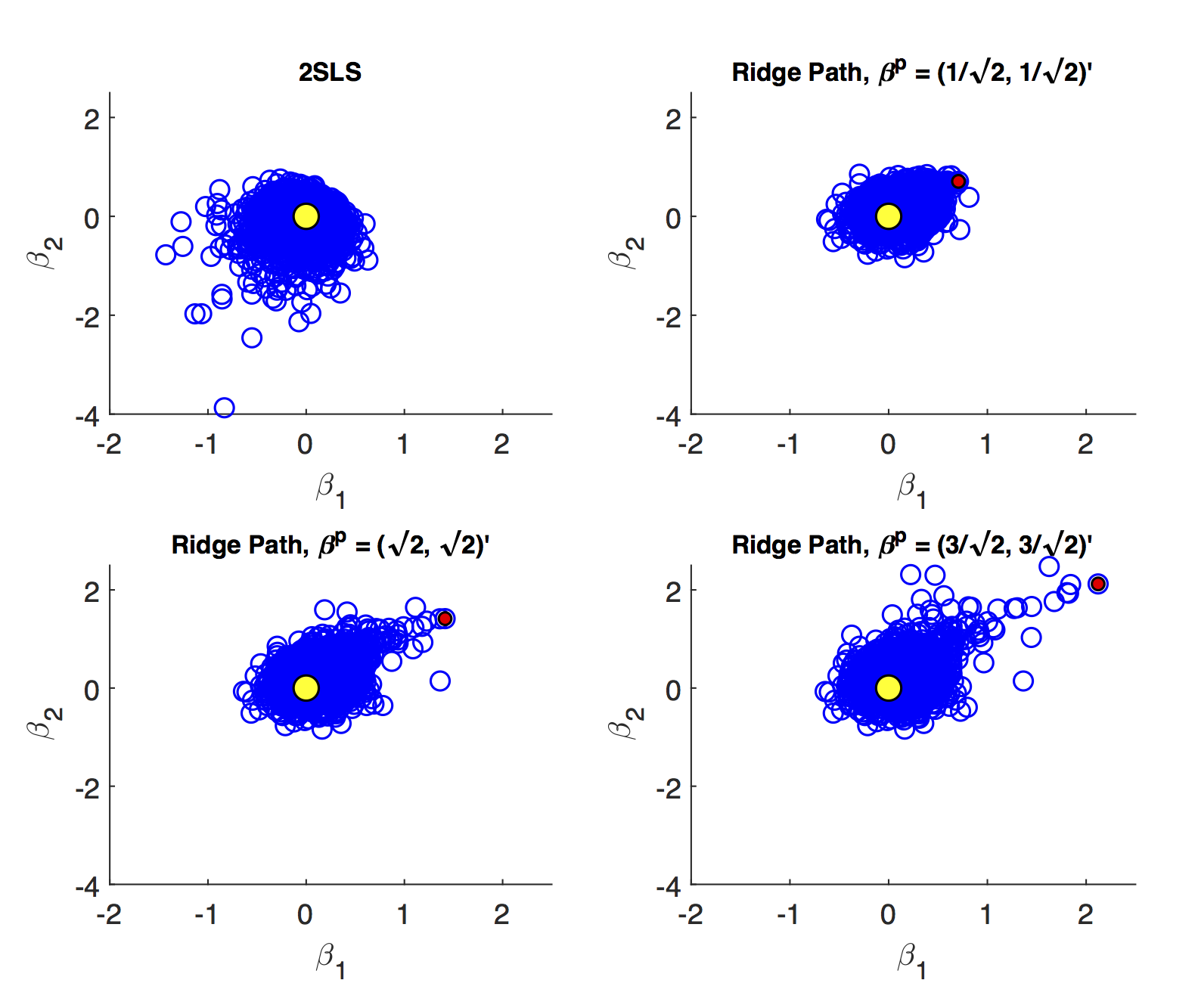}
\end{figure}

\begin{figure}[H]
\centering
\caption{Scatter plots of the estimates from 2SLS and ridge path estimator with different priors when precision is high ($\delta = 1$) and sample size is large ($n = 500$). Estimates, the true parameter value and prior values are represented by blue, yellow and red points respectively. The 2SLS estimates outperform the ridge path estimators which is demonstrated by marginally larger clouds which are slightly off-center from the true parameter values for the ridge path estimators. However, the ridge path estimator using different priors is still competitive. }\label{dlt100n500}
\includegraphics[width=0.95\linewidth]{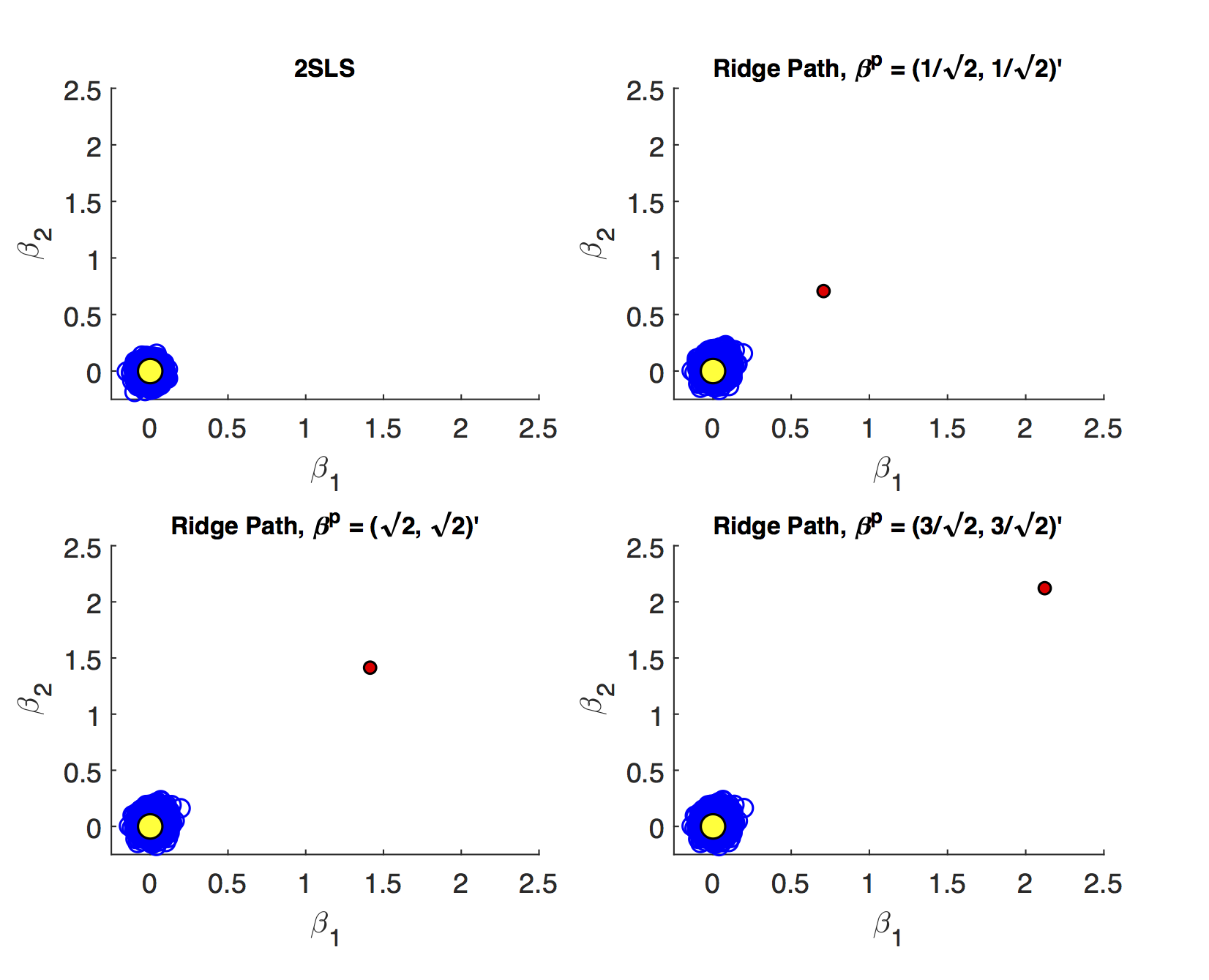}
\end{figure}

Figures \ref{dlt10n25} - \ref{dlt100n500} present scatter plots of the estimates from 2SLS and ridge path estimator with different priors for the following cases: a) low precision, small sample size;  b) low precision, large sample size;  c) high precision, small sample size; d) high precision, large sample size. These figures demonstrate the influence of the priors. The prior pulls the ridge path  estimates away from the population parameter values. For low precision models ($\delta = 0.1$), the variance associated with 2SLS estimates is larger than the ridge path estimates, even in larger sample sizes.  The ridge path estimator is biased towards the prior which is demonstrated by the estimates not being distributed symmetrically around the true value. On the other hand, for high precision models ($\delta = 1$) the variance reduction from 2SLS for the ridge path estimator is not as dramatic. In fact, while the variance reduction appears substantial for the prior value of  $\beta^p = (\frac{1}{\sqrt{2}} , \frac{1}{\sqrt{2}})'$, it is unclear at least visually if there is a reduction in variance for a poorly specified prior at $\beta^p = (\frac{3}{\sqrt{2}} , \frac{3}{\sqrt{2}})'$. In larger samples with high precision (Figure \ref{dlt100n500}) the 2SLS estimates outperform the ridge path estimators which is demonstrated by larger clouds which are slightly off-center from the true parameter values. However, ridge path estimators using different priors are still competitive and don't lead to a drastically worse performance (as a reference compare the performance of the 2SLS estimates to the ridge path estimates in Figure \ref{dlt10n25}). 

\begin{figure}[H]
\caption{This figure plots the histogram of estimated regularization parameter $\hat{\alpha}$ when $n = 10,000$ for all precision parameters and all priors considered in the simulations. The total number of simulations to generate each of these plots is $N = 1000$. As predicted by Theorem 1, the mass at $\hat{\alpha} = 0$ is approaching $50\%$ asymptotically. Distributions of $\hat{\alpha}$ values for all cases considered are presented in Table \ref{tabAlpha}.   }\label{fig: histAlpha}
\includegraphics[width=\linewidth]{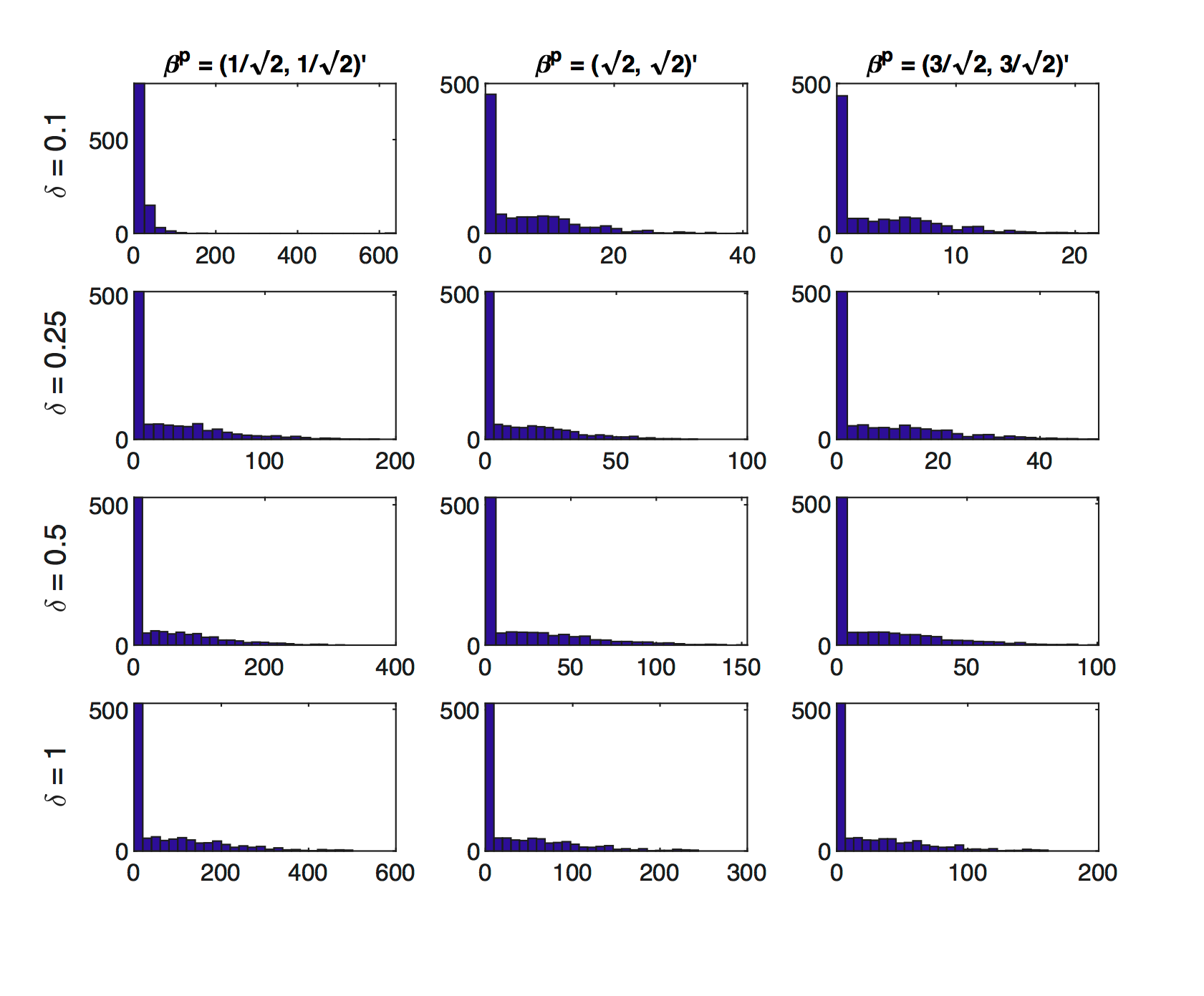}
\end{figure}

Table \ref{tabAlpha}, summarizes the distribution of the estimated regularization parameter $\hat{\alpha}$ for different precision levels, sample sizes and prior values. Recall Theorem 1 implies the asymptotic distribution will be a mixed distribution with some discrete mass at $\alpha = 0.$ Table \ref{tabAlpha} reports the proportion of cases which correspond to ``no regularization" ($\hat{\alpha} = 0$), ``infinite regularization" ($\hat{\alpha} = 10^7 \approx \infty$) and ``some regularization" ($\hat{\alpha} \in (0, 10^7)$). In all cases, there is a substantial mass of the distribution concentrated at $\hat{\alpha} = 0$. On the other hand we note that \textit{except} in the cases where the prior is located at the true parameter value, there is no mass concentrated at $\hat{\alpha} \approx \infty$. We see some interesting variations corresponding to different prior values. In low precision settings (particularly $\delta = 0.1$), keeping sample size fixed, as the prior moves away from the true value, the proportion of cases with ``no regularization" increases whereas the proportion of cases with ``some regularization" falls. Similarly for high precision settings (particularly $\delta = 1$), as  the sample size increases, the proportion of cases with ``no regularization" increases whereas the proportion of cases with ``some regularization" falls. In this table we also present results for large sample sizes of $n = 10,000$, which demonstrate that the mass at $\hat{\alpha} = 0$ approaches $50\%$ asymptotically, as predicted by Theorem 1. Distributions of $\hat{\alpha}$ for large sample sizes of $n = 10,000$ via histograms are presented in Figure \ref{fig: histAlpha}.  
%As this declines the precision of the model declines, for a fixed sample size. The condition number of the expectation of the gradient demonstrated these decline. 

Table \ref{tabSingular} presents summaries of the smallest singular value of the matrix\footnote{This corresponds to the estimate of $E\left[\frac{\partial g_i(\beta)}{\partial \beta'} \right]$ where $g_i(\beta) = (y_i-x_i\beta)z_i$.} $\left( \frac{-X'Z}{n} \right)$  for different values of $\delta$ and $n$.  The estimated asymptotic standard deviation is inversely related to the smallest singular value, or equivalently smaller singular values are associated with flatter objective functions at their minimum values.   As the precision parameter increases from $\delta = 0.1$ to $\delta = 1$, the mean of the smallest singular value increases.  As the sample size increases, the variance of the smallest singular values decreases.

%The general message from the simulations is that for low precision  models with small samples, the ridge path estimator is always superior to the 2SLS estimator. However, if the model has high precision and the sample size is large, then the 2SLS estimator is best. If the prior is close to the population parameter value then the ridge path estimator perform best even in the simulations with larger sample sizes. In models with low precision, the ridge path estimator usually leads to estimates with lower MSE values. When the model has low precision as well as when sample sizes are small, ridge path estimator leads to lower MSE. 

\section{Conclusion}\label{conclusion}

This paper addresses the problem of poor precision in linear IV estimation which occurs in samples where the objective function is flat in some dimension(s) at its minimum.  This results in imprecise estimates with high variances. S-sets and K-sets can be used to help address this problem, but without giving point estimates.  The main contribution of this paper is a method to obtain point estimates that can provide lower MSE than traditional 2SLS estimates when this problem occurs. The regularized point estimates presented are based on strong identification but address a practical gap in the literature where a point estimate is needed and hence the weak identification framework is inappropriate. 

A second contribution is the incorporation of a non-zero prior in the ridge path estimator. In the existing regularization literature within structural econometrics, the prior is typically fixed at the origin (following the machine learning literature). However, in structural econometric models, parameters  have meaningful interpretations.   Penalizing the discount factor and the risk aversion parameter towards zero is inappropriate and suggests the need to incorporate prior information. We show via simulations how a) the choice of prior affects the MSE and b) even poorly specified priors may outperform traditional 2SLS estimators in low precision or small sample size settings.  
 
A third contribution is the characterization of the nonstandard asymptotic distribution for the ridge path estimator.  This new approach incorporates the empirically selected tuning parameter into the asymptotic distribution.    

The chief benefit of these estimators is better small sample performance. Simulations demonstrate the trade-off of sample size and accuracy of the prior in determining the estimators small sample performance. The general message from the simulations is that for low precision  models, particularly with small samples, the ridge path estimator is  superior to the 2SLS estimator. If the model has high precision and the sample size is large, then the 2SLS estimator is best.  Fortunately, in these settings the ridge path estimator is competitive with the 2SLS estimator. If the prior is very close to, or at, the population parameter value then the ridge path estimator perform best in all simulations, including those with larger sample sizes. If the prior is away from the population parameter value, then the ridge path estimator's performance suffers; however even with a poorly defined prior the ridge path estimator may lead to lower MSE values, in low precision and small sample size settings, 

Open questions for future research include characterizing the behavior of the ridge path estimator with alternative types of models, such as weak instrument, or nearly weak instruments. Another important area for future research is extending the asymptotic proof technique to other empirical model selection rules such as k-fold cross validation.            

\pagebreak 

\begin{center}\begin{LARGE}\textbf{FIGURES AND TABLES}\end{LARGE}\end{center}

\begin{table}[H]
\centering

\caption{Estimates of $\hat{\beta}_1$ and $\hat{\beta}_2$ using 2SLS and ridge path estimator for $\beta^p = (\frac{1}{\sqrt{2}}, \frac{1}{\sqrt{2}})'$. The ridge path estimator leads to smaller combined MSE (highlighted in bold) when precision is low  ($\delta = 0.10$).  This drop in MSE is driven primarily by large reductions in standard deviations of the estimates. The 2SLS estimator leads to smaller combined MSE when precision is high ($\delta = 1.00$). For intermediate precision models the ridge path estimator leads to smaller combined MSE in small samples. } \label{tabBeta1}
 \begin{tabular}{ |l | l | l | l | l | l | l | l | l | l | }

 \hline
            &     &   &\multicolumn{3}{c|}{$\hat{\beta}_1$} &\multicolumn{3}{c|}{$\hat{\beta}_2$} &  {$(\hat{\beta}_1, \hat{\beta}_2)$} \\
            
            \cline{4-10}
{$\delta$} & {$n$}& {Estimator} & {Bias} & {SD} & {MSE} & {Bias} & {SD} & {MSE} & {MSE} \\
\hline
0.10 & 25 & 2SLS & 0.013 & 0.231 & 0.053 & 0.630 & 1.520 & 2.708 & 2.762 \\ 
 &  & Ridge Path & 0.146 & 0.121 & 0.036 & 0.685 & 0.248 & 0.531 & \textbf{0.567} \\ \cline{2-10} 
 & 50 & 2SLS & 0.006 & 0.189 & 0.036 & 0.546 & 1.427 & 2.334 & 2.370 \\ 
 &  & Ridge Path & 0.091 & 0.085 & 0.016 & 0.664 & 0.245 & 0.501 & \textbf{0.516} \\ \cline{2-10} 
 & 250 & 2SLS & -0.000 & 0.081 & 0.007 & 0.202 & 1.512 & 2.327 & 2.333 \\ 
 &  & Ridge Path & 0.032 & 0.041 & 0.003 & 0.560 & 0.256 & 0.380 & \textbf{0.382} \\ \cline{2-10}  
 & 500 & 2SLS & -0.000 & 0.041 & 0.002 & 0.060 & 0.762 & 0.584 & 0.585 \\ 
 &  & Ridge Path & 0.020 & 0.031 & 0.001 & 0.479 & 0.253 & 0.294 & \textbf{0.295} \\ \hline 
0.25 & 25 & 2SLS & 0.008 & 0.216 & 0.047 & 0.325 & 1.158 & 1.446 & 1.493 \\ 
 &  & Ridge Path & 0.149 & 0.129 & 0.039 & 0.599 & 0.250 & 0.422 & \textbf{0.461} \\ \cline{2-10}  
 & 50 & 2SLS & 0.002 & 0.148 & 0.022 & 0.176 & 1.098 & 1.236 & 1.258 \\ 
 &  & Ridge Path & 0.092 & 0.092 & 0.017 & 0.529 & 0.251 & 0.343 & \textbf{0.360} \\ \cline{2-10}  
 & 250 & 2SLS & -0.001 & 0.047 & 0.002 & -0.002 & 0.298 & 0.089 & \textbf{0.091} \\ 
 &  & Ridge Path & 0.025 & 0.046 & 0.003 & 0.292 & 0.222 & 0.135 & 0.137 \\ \cline{2-10}  
 & 500 & 2SLS & -0.000 & 0.032 & 0.001 & -0.000 & 0.188 & 0.035 & \textbf{0.036} \\ 
 &  & Ridge Path & 0.013 & 0.033 & 0.001 & 0.204 & 0.189 & 0.077 & 0.079 \\ \hline 
 0.50 & 25 & 2SLS & 0.002 & 0.199 & 0.040 & 0.053 & 0.747 & 0.561 & 0.600 \\ 
 &  & Ridge Path & 0.148 & 0.143 & 0.043 & 0.425 & 0.248 & 0.242 & \textbf{0.284} \\ \cline{2-10} 
 & 50 & 2SLS & 0.000 & 0.112 & 0.013 & 0.005 & 0.402 & 0.162 & 0.174 \\ 
 &  & Ridge Path & 0.085 & 0.100 & 0.017 & 0.318 & 0.220 & 0.150 & \textbf{0.167} \\ \cline{2-10}  
 & 250 & 2SLS & -0.001 & 0.045 & 0.002 & -0.001 & 0.131 & 0.017 & \textbf{0.019} \\ 
 &  & Ridge Path & 0.023 & 0.048 & 0.003 & 0.130 & 0.139 & 0.036 & 0.039 \\ \cline{2-10}  
 & 500 & 2SLS & 0.000 & 0.032 & 0.001 & 0.000 & 0.090 & 0.008 & \textbf{0.009} \\ 
 &  & Ridge Path & 0.013 & 0.035 & 0.001 & 0.087 & 0.103 & 0.018 & 0.020 \\ \hline 
 1.00 & 25 & 2SLS & -0.002 & 0.163 & 0.026 & -0.003 & 0.244 & 0.060 & \textbf{0.086} \\ 
 &  & Ridge Path & 0.139 & 0.153 & 0.043 & 0.213 & 0.195 & 0.083 & 0.126 \\ \cline{2-10} 
 & 50 & 2SLS & 0.000 & 0.106 & 0.011 & 0.001 & 0.153 & 0.023 & \textbf{0.035} \\ 
 &  & Ridge Path & 0.082 & 0.105 & 0.018 & 0.141 & 0.147 & 0.041 & 0.059 \\ \cline{2-10}  
 & 250 & 2SLS & -0.001 & 0.045 & 0.002 & -0.000 & 0.064 & 0.004 & \textbf{0.006} \\ 
 &  & Ridge Path & 0.028 & 0.050 & 0.003 & 0.053 & 0.073 & 0.008 & 0.011 \\  
 & 500 & 2SLS & 0.000 & 0.032 & 0.001 & 0.000 & 0.045 & 0.002 & \textbf{0.003} \\ 
 &  & Ridge Path & 0.019 & 0.036 & 0.002 & 0.036 & 0.053 & 0.004 & 0.006 \\
\bottomrule

\end{tabular}
\end{table}

\begin{table}
\centering

\caption{Estimates of $\hat{\beta}_1$ and $\hat{\beta}_2$ using 2SLS and ridge path estimator for $\beta^p = (\frac{3}{\sqrt{2}}, \frac{3}{\sqrt{2}})'$. The prior is 3 standard deviations away from the true parameter value, the ridge path estimator outperforms the 2SLS estimator in terms of  MSE values in a number of cases. In particular, in small samples and low precision settings, the ridge path estimator leads to smaller MSE values. }\label{tabBeta3}
 \begin{tabular}{ |l | l | l | l | l | l | l | l | l | l | }
 \hline
            &     &   &\multicolumn{3}{c|}{$\hat{\beta}_1$} &\multicolumn{3}{c|}{$\hat{\beta}_2$} &  {$(\hat{\beta}_1, \hat{\beta}_2)$} \\
            
            \cline{4-10}
{$\delta$} & {$n$}& {Estimator} & {Bias} & {SD} & {MSE} & {Bias} & {SD} & {MSE} & {MSE} \\

\hline
0.10 & 25 & 2SLS & 0.013 & 0.232 & 0.054 & 0.628 & 1.506 & 2.662 & 2.716 \\ 
 &  & Ridge Path & 0.131 & 0.193 & 0.054 & 1.056 & 0.536 & 1.401 & \textbf{1.456} \\ \cline{2-10}
 & 50 & 2SLS & 0.006 & 0.190 & 0.036 & 0.546 & 1.435 & 2.356 & 2.392 \\ 
 &  & Ridge Path & 0.072 & 0.130 & 0.022 & 1.024 & 0.547 & 1.348 & \textbf{1.370} \\  \cline{2-10}
 & 250 & 2SLS & -0.000 & 0.081 & 0.007 & 0.203 & 1.509 & 2.318 & 2.324 \\ 
 &  & Ridge Path & 0.017 & 0.051 & 0.003 & 0.800 & 0.527 & 0.917 & \textbf{0.920} \\  \cline{2-10}
 & 500 & 2SLS & -0.000 & 0.041 & 0.002 & 0.062 & 0.757 & 0.576 & \textbf{0.578} \\ 
 &  & Ridge Path & 0.008 & 0.034 & 0.001 & 0.623 & 0.465 & 0.604 & 0.605 \\ \hline
0.25 & 25 & 2SLS & 0.008 & 0.217 & 0.047 & 0.324 & 1.167 & 1.466 & 1.513 \\ 
 &  & Ridge Path & 0.128 & 0.196 & 0.055 & 0.887 & 0.525 & 1.062 & \textbf{1.116} \\  \cline{2-10}
 & 50 & 2SLS & 0.002 & 0.148 & 0.022 & 0.176 & 1.088 & 1.215 & 1.237 \\ 
 &  & Ridge Path & 0.066 & 0.121 & 0.019 & 0.749 & 0.499 & 0.810 & \textbf{0.829} \\  \cline{2-10}
& 250 & 2SLS & -0.001 & 0.047 & 0.002 & -0.002 & 0.298 & 0.089 & \textbf{0.091} \\ 
 &  & Ridge Path & 0.013 & 0.045 & 0.002 & 0.325 & 0.276 & 0.182 & 0.184 \\  \cline{2-10}
 & 500 & 2SLS & 0.000 & 0.032 & 0.001 & 0.000 & 0.188 & 0.035 & \textbf{0.036} \\ 
 &  & Ridge Path & 0.008 & 0.033 & 0.001 & 0.208 & 0.197 & 0.082 & 0.083 \\  \hline
0.50 & 25 & 2SLS & 0.002 & 0.199 & 0.040 & 0.052 & 0.750 & 0.565 & 0.605 \\ 
 &  & Ridge Path & 0.122 & 0.180 & 0.047 & 0.555 & 0.429 & 0.491 & \textbf{0.539} \\  \cline{2-10}
 & 50 & 2SLS & -0.000 & 0.113 & 0.013 & 0.005 & 0.402 & 0.162 & \textbf{0.174} \\ 
 &  & Ridge Path & 0.063 & 0.108 & 0.016 & 0.374 & 0.307 & 0.234 & 0.250 \\  \cline{2-10}
 & 250 & 2SLS & -0.001 & 0.046 & 0.002 & -0.001 & 0.130 & 0.017 & \textbf{0.019} \\ 
 &  & Ridge Path & 0.018 & 0.048 & 0.003 & 0.132 & 0.142 & 0.038 & 0.040 \\  \cline{2-10}
 & 500 & 2SLS & 0.000 & 0.032 & 0.001 & 0.000 & 0.091 & 0.008 & \textbf{0.009} \\ 
 &  & Ridge Path & 0.011 & 0.035 & 0.001 & 0.088 & 0.103 & 0.018 & 0.020 \\ \hline
1.00 & 25 & 2SLS & -0.002 & 0.162 & 0.026 & -0.004 & 0.244 & 0.059 & \textbf{0.086} \\ 
 &  & Ridge Path & 0.126 & 0.169 & 0.044 & 0.226 & 0.236 & 0.107 & 0.151 \\  \cline{2-10}
 & 50 & 2SLS & 0.000 & 0.106 & 0.011 & 0.001 & 0.153 & 0.023 & \textbf{0.035} \\ 
 &  & Ridge Path & 0.075 & 0.109 & 0.018 & 0.145 & 0.158 & 0.046 & 0.063 \\  \cline{2-10}
 & 250 & 2SLS & -0.001 & 0.045 & 0.002 & -0.000 & 0.064 & 0.004 & \textbf{0.006} \\ 
 &  & Ridge Path & 0.026 & 0.051 & 0.003 & 0.053 & 0.074 & 0.008 & 0.012 \\  \cline{2-10}
 & 500 & 2SLS & -0.000 & 0.032 & 0.001 & -0.000 & 0.045 & 0.002 & \textbf{0.003} \\ 
 &  & Ridge Path & 0.018 & 0.037 & 0.002 & 0.036 & 0.053 & 0.004 & 0.006 \\ \hline
\bottomrule
\end{tabular}

\end{table}

\begin{landscape}
\begin{table}[H]
\centering

\caption{Distribution of Regularization Parameter $\hat{\alpha}$. The mixed distribution associated with the finite samples is in agreement with the nonstandard asymptotic distribution given in Theorem 1.  The proportion of cases with ``no regularization" ($\hat{\alpha} = 0$), ``some regularization" ($\hat{\alpha} \in (0, 10^7)$) and ``infinite regularization" ($\hat{\alpha} = 10^7 \approx \infty$) are presented . For all cases, there is a substantial mass of the distribution concentrated at $\hat{\alpha} = 0$. On the other hand, there is no mass concentrated at $\hat{\alpha} \approx \infty$ \textit{except} in very small samples of $n = 25$. As predicted by Theorem 1, the mass at $\hat{\alpha} = 0$ is approaching $50\%$ asymptotically.  Histograms for the large sample cases of $n = 10,000$ are presented in Figure \ref{fig: histAlpha}.  }\label{tabAlpha}

 \begin{tabular}{ 
c|c| c c c | c c c |c c c 
 }
 \toprule
  & & \multicolumn{3}{|c|}{ $\beta^p = (1/\sqrt{2}, 1/\sqrt{2})'$ }&\multicolumn{3}{c|}{ $\beta^p = (\sqrt{2}, \sqrt{2})'$ }&\multicolumn{3}{c}{ $\beta^p = (3/\sqrt{2}, 3/\sqrt{2})'$ } \\
  \cline{3-11}
$\delta$ & $n$ &  $\hat{\alpha} = 0$ &  $\hat{\alpha} \in (0, 10^7)$& $\hat{\alpha} = 10^7 \approx \infty$ & $\hat{\alpha} = 0$ &  $\hat{\alpha} \in (0, 10^7)$& $\hat{\alpha} = 10^7$ & $\hat{\alpha} = 0$ & $\hat{\alpha} \in (0, 10^7)$  & $\hat{\alpha} = 10^7$\\ 
\hline

0.01 & 25 & 0.164 & 0.834 & 0.003 & 0.262 & 0.738 & 0.001 & 0.339 & 0.661 & 0.000 \\ 
 & 50 & 0.166 & 0.834 & 0.000 & 0.275 & 0.725 & 0.000 & 0.354 & 0.646 & 0.000 \\ 
 & 250 & 0.190 & 0.810 & 0.000 & 0.281 & 0.719 & 0.000 & 0.319 & 0.681 & 0.000 \\ 
 & 500 & 0.220 & 0.780 & 0.000 & 0.285 & 0.715 & 0.000 & 0.302 & 0.698 & 0.000 \\ 
 & 10000 & 0.413 & 0.587 & 0.000 & 0.411 & 0.589 & 0.000 & 0.411 & 0.589 & 0.000 \\ \hline 
0.25 & 25 & 0.176 & 0.822 & 0.002 & 0.262 & 0.737 & 0.001 & 0.315 & 0.684 & 0.001 \\ 
& 50 & 0.184 & 0.816 & 0.000 & 0.263 & 0.737 & 0.000 & 0.299 & 0.701 & 0.000 \\ 
& 250 & 0.293 & 0.707 & 0.000 & 0.309 & 0.691 & 0.000 & 0.314 & 0.686 & 0.000 \\ 
& 500 & 0.346 & 0.654 & 0.000 & 0.354 & 0.646 & 0.000 & 0.354 & 0.646 & 0.000 \\ 
& 10000 & 0.461 & 0.539 & 0.000 & 0.465 & 0.535 & 0.000 & 0.465 & 0.535 & 0.000 \\ \hline 
0.50 & 25 & 0.216 & 0.780 & 0.004 & 0.262 & 0.737 & 0.001 & 0.284 & 0.716 & 0.000 \\  
& 50 & 0.255 & 0.745 & 0.000 & 0.284 & 0.716 & 0.000 & 0.294 & 0.706 & 0.000 \\  
& 250 & 0.369 & 0.631 & 0.000 & 0.374 & 0.626 & 0.000 & 0.376 & 0.624 & 0.000 \\ 
& 500 & 0.412 & 0.588 & 0.000 & 0.415 & 0.585 & 0.000 & 0.417 & 0.583 & 0.000 \\ 
& 10000 & 0.463 & 0.537 & 0.000 & 0.467 & 0.533 & 0.000 & 0.466 & 0.534 & 0.000 \\ \hline 
1.00& 25 & 0.287 & 0.708 & 0.005 & 0.310 & 0.689 & 0.001 & 0.318 & 0.681 & 0.000 \\ 
& 50 & 0.333 & 0.667 & 0.000 & 0.346 & 0.654 & 0.000 & 0.351 & 0.649 & 0.000 \\ 
& 250 & 0.413 & 0.587 & 0.000 & 0.418 & 0.582 & 0.000 & 0.419 & 0.581 & 0.000 \\ 
& 500 & 0.439 & 0.561 & 0.000 & 0.443 & 0.557 & 0.000 & 0.442 & 0.558 & 0.000 \\ 
& 10000 & 0.474 & 0.526 & 0.000 & 0.478 & 0.522 & 0.000 & 0.477 & 0.523 & 0.000 \\ 
\bottomrule
\end{tabular}
\end{table}

\end{landscape}

\pagebreak

\begin{table}[H]
\centering
\caption{Summary statistics of the smallest singular value for the matrix $ \left(- \frac{X'Z}{n} \right)$ corresponding to different precision parameter values $\delta$ and sample sizes $n$, using $10,000$ samples each. As the precision parameters increase from $\delta = 0.1$ to $\delta = 1$, the mean of the smallest singular value increases. As sample sizes increase from $n = 25$ to $n = 10,000$, the spread in the smallest singular value decreases.  }\label{tabSingular}
 \begin{tabular}{ |c | l | c | c | c | c | c | }
\hline
$\delta$ & $n$ & Mean & Std Dev & $1^{st}$ Quartile & Median & $3^{rd}$ Quartile \\\hline
0.10 & 25 & 0.25 & 0.14 & 0.14 & 0.23 & 0.33 \\ 
 & 50 & 0.19 & 0.10 & 0.12 & 0.18 & 0.26 \\ 
 & 250 & 0.12 & 0.05 & 0.08 & 0.12 & 0.16 \\ 
 & 500 & 0.11 & 0.04 & 0.08 & 0.11 & 0.14 \\ 
 & 2500 & 0.10 & 0.02 & 0.09 & 0.10 & 0.12 \\ 
 & 5000 & 0.10 & 0.01 & 0.09 & 0.10 & 0.11 \\ 
 & 10000 & 0.10 & 0.01 & 0.09 & 0.10 & 0.11 \\ \hline 
0.25 & 25 & 0.32 & 0.17 & 0.19 & 0.30 & 0.42 \\ 
 & 50 & 0.28 & 0.13 & 0.19 & 0.27 & 0.37 \\  
 & 250 & 0.26 & 0.07 & 0.21 & 0.26 & 0.30 \\  
 & 500 & 0.25 & 0.05 & 0.22 & 0.25 & 0.29 \\ 
 & 2500 & 0.25 & 0.02 & 0.24 & 0.25 & 0.26 \\ 
 & 5000 & 0.25 & 0.01 & 0.24 & 0.25 & 0.26 \\ 
 & 10000 & 0.25 & 0.01 & 0.24 & 0.25 & 0.26 \\ \hline 
0.50 & 25 & 0.50 & 0.21 & 0.35 & 0.48 & 0.63 \\ 
 & 50 & 0.50 & 0.17 & 0.39 & 0.49 & 0.61 \\  
 & 250 & 0.50 & 0.08 & 0.45 & 0.50 & 0.55 \\  
 & 500 & 0.50 & 0.05 & 0.46 & 0.50 & 0.54 \\ 
 & 2500 & 0.50 & 0.02 & 0.48 & 0.50 & 0.52 \\ 
 & 5000 & 0.50 & 0.02 & 0.49 & 0.50 & 0.51 \\  
 & 10000 & 0.50 & 0.01 & 0.49 & 0.50 & 0.51 \\ \hline 
1.00 & 25 & 0.86 & 0.27 & 0.67 & 0.85 & 1.03 \\ 
 & 50 & 0.92 & 0.21 & 0.77 & 0.91 & 1.05 \\ 
 & 250 & 0.98 & 0.10 & 0.91 & 0.98 & 1.05 \\ 
 & 500 & 0.99 & 0.08 & 0.94 & 0.99 & 1.04 \\  
 & 2500 & 1.00 & 0.03 & 0.98 & 1.00 & 1.02 \\  
 & 5000 & 1.00 & 0.02 & 0.98 & 1.00 & 1.02 \\ 
 & 10000 & 1.00 & 0.02 & 0.99 & 1.00 & 1.01 \\ \hline 
\hline
\end{tabular}
\end{table}

\bigskip

%%% BIBLIOGRAPHY
\bibliographystyle{plain}
\nocite{*}
\bibliography{biblioMar2019}

\newpage

\appendix

\begin{center}\begin{LARGE}\textbf{APPENDIX}\end{LARGE}\end{center}
%\begin{small}

\section{Proof of Lemma 1}

The objective function that determines the optimal tuning parameter is given in equation (\ref{alpha_equation}).    As the sample size grows the objective function uniformly converges to   a deterministic function that takes a unique local minimum at $\alpha = 0.$   The parameter space is bounded and the law of large numbers implies    
$$
\lim_{n \rightarrow \infty} Q_{n(1-\tau)}(\alpha) = \frac{1}{2}
\left(   \beta_0 -  \beta^p  \right)'  \left( \frac{ \Gamma_0' R_z \Gamma_0 }{\alpha} + I_k  \right)^{-1}\Gamma_0' R_z \Gamma_0  
 \left( \frac{ \Gamma_0' R_z \Gamma_0}{\alpha} +  I_k  \right)^{-1} \left( \beta_0  -\beta^p  \right)    
$$ 
which is minimized at $\alpha = 0$.  Hence $\alpha_0 = 0.$ When $\alpha = 0 $ then $\hat{\beta}_{IV, \tau n}(0) \rightarrow \beta_0$. 

The root-$n$ consistency of $\hat{\alpha}$ follows from the standard approach of Lemma 5.4 in Ichimura (1993).  The needed results are that 
$
\frac{d Q_{n(1-\tau)}(\alpha_0)}{d \alpha}
$
satisfies a CLT and $ \frac{d^2 Q_{n(1-\tau)}(\alpha)}{d \alpha^2} $ is continuous (from the right hand side) at $\alpha_0$ and $ \frac{d^2 Q_{n(1-\tau)}(\alpha_0)}{d \alpha^2} $ limits to a positive value.   These derivatives reduce to the derivatives of 
$
\hat{\beta}_{IV, \tau n}(\alpha) =  \left( 
\frac{X_{\tau n}'
P_{Z_{\tau n}} 
 X_{\tau n}}{ [ \tau n ] } 
+ \alpha I \right)^{-1} \left( \frac{X_{\tau n}'
P_{Z_{\tau n}} 
 Y_{\tau n} }{ [ \tau n ] }
  + \alpha \beta^p \right)
$
wrt $\alpha$.
The first derivative is 
\begin{eqnarray*}
\frac{d \hat{\beta}_{IV, \tau n}(\alpha)}{d \alpha}
& = &\left(  
\frac{X_{\tau n}'
P_{Z_{\tau n}} 
 X_{\tau n}}{ [ \tau n ] }  + \alpha I_k\right)^{-1}  \beta^p 
- \left( 
\frac{X_{\tau n}'
P_{Z_{\tau n}} 
 X_{\tau n}}{ [ \tau n ] }  + \alpha I_k\right)^{-2}\left( \frac{X_{\tau n}'
P_{Z_{\tau n}} 
 Y_{\tau n} }{ [ \tau n ] }+  \alpha \beta^p \right)
\\
&\ = & \left(  
\frac{X_{\tau n}'
P_{Z_{\tau n}} 
 X_{\tau n}}{ [ \tau n ] }  + \alpha I_k\right)^{-1} \left(  \beta^p - \hat{\beta}_{IV, \tau n}(\alpha) 
 \right).
\end{eqnarray*}
The second derivative is 
\begin{eqnarray*}
\frac{d^{2} \hat{\beta}_{IV, \tau n}(\alpha)}{d \alpha^{2}}
& = &-\left( \frac{X_{\tau n}'
P_{Z_{\tau n}} 
 X_{\tau n}}{ [ \tau n ] }  + \alpha I_k\right)^{-1}  \frac{d \hat{\beta}_{IV, \tau n}(\alpha)}{d \alpha} 
- \left( \frac{X_{\tau n}'
P_{Z_{\tau n}} 
 X_{\tau n}}{ [ \tau n ] }  + \alpha I_k\right)^{-2}\left(  \beta^p - \hat{\beta}_{IV, \tau n}(\alpha) 
 \right)
\\
& = & -\left( \frac{X_{\tau n}'
P_{Z_{\tau n}} 
 X_{\tau n}}{ [ \tau n ] }  + \alpha I_k\right)^{-2} 
 \left(  \beta^p - \hat{\beta}_{IV, \tau n}(\alpha) 
 \right) 
- \left( \frac{X_{\tau n}'
P_{Z_{\tau n}} 
 X_{\tau n}}{ [ \tau n ] }  + \alpha I_k\right)^{-2}\left(  \beta^p - \hat{\beta}_{IV, \tau n}(\alpha) 
 \right)
\\
& = & -2\left( \frac{X_{\tau n}'
P_{Z_{\tau n}} 
 X_{\tau n}}{ [ \tau n ] }  + \alpha I_k\right)^{-2} 
 \left(  \beta^p - \hat{\beta}_{IV, \tau n}(\alpha)
 \right). 
\end{eqnarray*}
Now determine the derivatives of $Q_{n(1 - \tau)}(\alpha)$.  The first derivative is
\begin{eqnarray*}
\frac{d Q_{n(1-\tau)}(\alpha)}{d \alpha } & = & 
\frac{-1}{( n - [\tau n] ) } \left( Y_{n(1-\tau)} - X_{n(1-\tau)} \hat{\beta}_{IV, \tau n}(\alpha) \right)'
 P_{Z_{ n(1-\tau)}}
  X_{n(1-\tau)} 
\frac{ d \hat{\beta}_{IV, \tau n}(\alpha) }{ d \alpha }   
\\
& = &  \frac{-1}{( n - [\tau n] ) } \left( Y_{n(1-\tau)} - X_{n(1-\tau)} \hat{\beta}_{IV, \tau n}(\alpha) \right)'
 P_{Z_{ n(1-\tau)}}
  X_{n(1-\tau)}  \left(  
\frac{X_{\tau n}'
P_{Z_{\tau n}} 
 X_{\tau n}}{ [ \tau n ] }  + \alpha I_k\right)^{-1} \left(  \beta^p - \hat{\beta}_{IV, \tau n}(\alpha) 
 \right).
\end{eqnarray*}
Evaluate at $\alpha_0 = 0$  
\begin{eqnarray*}
\frac{d Q_{n(1-\tau)}(0)}{d \alpha }
& = & \frac{-1}{( n - [\tau n] ) }  \left( Y_{n(1-\tau)} - X_{n(1-\tau)} \hat{\beta}_{IV, \tau n}(0) \right)'
 P_{Z_{ n(1-\tau)}}
  X_{n(1-\tau)}  \left(  
\frac{X_{\tau n}'
P_{Z_{\tau n}} 
 X_{\tau n}}{ [ \tau n ] }  \right)^{-1} \left(  \beta^p - \hat{\beta}_{IV, \tau n}(0 ) 
 \right)
\\
& = &  \frac{-1}{( n - [\tau n] ) }  \left( (Y_{n(1-\tau)} - X_{n(1-\tau)} \beta_0) - X_{n(1-\tau)} ( \hat{\beta}_{IV, \tau n}(0) -\beta_0)
\right)' \nonumber
 \\ 
& & \times  P_{Z_{ n(1-\tau)}}
  X_{n(1-\tau)}  \left(  
\frac{X_{\tau n}'
P_{Z_{\tau n}} 
 X_{\tau n}}{ [ \tau n ] }  \right)^{-1} \left(  \beta^p - \beta_0  
  -( \hat{\beta}_{IV, \tau n}(0) -\beta_0)\right) 
\\
& = &  \frac{-1}{( n - [\tau n] ) }  \left( \varepsilon_{n(1-\tau)}' - 
\frac{ \varepsilon_{\tau n}'
P_{Z_{\tau n}} 
 X_{\tau n}}{ [ \tau n ] } \left( 
\frac{X_{\tau n}'
P_{Z_{\tau n}} 
 X_{\tau n}}{ [ \tau n ] } \right)^{-1} X_{n(1-\tau)}' \right) 
\nonumber
\\
 & & \times   P_{Z_{ n(1-\tau)}}
  X_{n(1-\tau)}  \left(  
\frac{X_{\tau n}'
P_{Z_{\tau n}} 
 X_{\tau n}}{ [ \tau n ] }  \right)^{-1} \left(  \beta^p - \beta_0  
- 
 \left( 
\frac{X_{\tau n}'
P_{Z_{\tau n}} 
 X_{\tau n}}{ [ \tau n ] } \right)^{-1}
\frac{ X_{\tau n}'
P_{Z_{\tau n}} 
 \varepsilon_{\tau n}}{ [ \tau n ] } 
 \right). 
\end{eqnarray*}
The CLT applies to the $\varepsilon_{n(1-\tau)}' Z_{n(1-\tau)}$ and $\varepsilon_{\tau n}' Z_{ \tau n}$ terms.  The others converge by LLN. 
Hence
\begin{eqnarray*}
\sqrt{(n - [\tau n] )} \frac{d Q_{n(1-\tau)}(0)}{d \alpha }
& = & \frac{-1}{\sqrt{( n - [\tau n] )} }  \left( \varepsilon_{n(1-\tau)}' - 
\frac{ \varepsilon_{\tau n}'
P_{Z_{\tau n}} 
 X_{\tau n}}{ [ \tau n ] } \left( 
\frac{X_{\tau n}'
P_{Z_{\tau n}} 
 X_{\tau n}}{ [ \tau n ] } \right)^{-1} X_{n(1-\tau)}' \right) \nonumber
\\
 & & \times   P_{Z_{ n(1-\tau)}}
  X_{n(1-\tau)}  \left(  
\frac{X_{\tau n}'
P_{Z_{\tau n}} 
 X_{\tau n}}{ [ \tau n ] }  \right)^{-1} \left(  \beta^p - \beta_0  
 \right) + o_p(1). 
\end{eqnarray*}

The second derivative is %
\begin{eqnarray*}
& & 
\frac{d^{2} Q_{n(1-\tau)}(\alpha)}{d \alpha^{2} }
\\
& = & 
\frac{-1}{( n - [\tau n] ) } \left( Y_{n(1-\tau)} - X_{n(1-\tau)} \hat{\beta}_{IV, \tau n}(\alpha) \right)'
 P_{Z_{ n(1-\tau)}}
  X_{n(1-\tau)} 
\frac{ d^2 \hat{\beta}_{IV, \tau n}(\alpha) }{ d \alpha^2 }   
\\
& & + 
\frac{1}{( n - [\tau n] ) } \left(  X_{n(1-\tau)} \frac{ d \hat{\beta}_{IV, \tau n}(\alpha) }{ d \alpha } \right)'
 P_{Z_{ n(1-\tau)}}
  X_{n(1-\tau)} 
\frac{ d \hat{\beta}_{IV, \tau n}(\alpha) }{ d \alpha }   
\\
& = & \frac{2}{( n - [\tau n] ) }\left( Y_{n(1-\tau)} - X_{n(1-\tau)} \hat{\beta}_{IV, \tau n}(\alpha) \right)'
%\\
%%
%& & \hspace{.5in}
 P_{Z_{ n(1-\tau)}}
  X_{n(1-\tau)}  \left( \frac{X_{\tau n}'
P_{Z_{\tau n}} 
 X_{\tau n}}{ [ \tau n ] }  + \alpha I_k\right)^{-2} 
 \left(  \beta^p - \hat{\beta}_{IV, \tau n}(\alpha)
 \right) \\
& & + \frac{1}{( n - [\tau n] ) } 
 \left(  \beta^p - \hat{\beta}_{IV, \tau n}(\alpha) \right)'
\left(  
\frac{X_{\tau n}'
P_{Z_{\tau n}} 
 X_{\tau n}}{ [ \tau n ] }  + \alpha I_k\right)^{-1}
%\\
%%
%& & 
%\hspace{.5in}
%\times  
 X_{n(1-\tau)}'
\\
& & \hspace{2in} \times 
P_{Z_{ n(1-\tau)}}
  X_{n(1-\tau)}  \left(  
\frac{X_{\tau n}'
P_{Z_{\tau n}} 
 X_{\tau n}}{ [ \tau n ] }  + \alpha I_k\right)^{-1} \left(  \beta^p - \hat{\beta}_{IV, \tau n}(\alpha) 
 \right).
\end{eqnarray*}
This is a bounded continuous function.  Now evaluate at $\alpha_{0} = 0$

\begin{eqnarray*}
\frac{d^{2} Q_{n(1-\tau)}(0)}{d \alpha^{2} }
& = & \frac{2}{( n - [\tau n] ) }\left( (Y_{n(1-\tau)} - X_{n(1-\tau)} \beta_0) - X_{n(1-\tau)} ( \hat{\beta}_{IV, \tau n}(0) -\beta_0)
\right)'   \nonumber 
\\
& & \hspace{.5in}
\times P_{Z_{ n(1-\tau)}}
  X_{n(1-\tau)}  \left( \frac{X_{\tau n}'
P_{Z_{\tau n}} 
 X_{\tau n}}{ [ \tau n ] } \right)^{-2} 
 \left(  \beta^p -  \beta_0 - (\hat{\beta}_{IV, \tau n}(0) - \beta_0)
 \right)   \nonumber \\
& & + \frac{1}{( n - [\tau n] ) } 
 \left(  \beta^p -  \beta_0 - (\hat{\beta}_{IV, \tau n}(0) - \beta_0)
 \right)'
\left(  
\frac{X_{\tau n}'
P_{Z_{\tau n}} 
 X_{\tau n}}{ [ \tau n ] }  \right)^{-1}  \nonumber 
\\
& & 
\hspace{.5in}
\times  
 X_{n(1-\tau)}'
 P_{Z_{ n(1-\tau)}}
  X_{n(1-\tau)}  \left(  
\frac{X_{\tau n}'
P_{Z_{\tau n}} 
 X_{\tau n}}{ [ \tau n ] }  \right)^{-1} \left(  \beta^p -  \beta_0 - (\hat{\beta}_{IV, \tau n}(0) - \beta_0)
 \right)
\\
& = & \frac{2}{( n - [\tau n] ) }\left( \varepsilon_{n(1-\tau)} - X_{n(1-\tau)} ( \hat{\beta}_{IV, \tau n}(0) -\beta_0)
\right)'  \nonumber 
\\
& & \hspace{.5in}
\times P_{Z_{ n(1-\tau)}}
  X_{n(1-\tau)}  \left( \frac{X_{\tau n}'
P_{Z_{\tau n}} 
 X_{\tau n}}{ [ \tau n ] } \right)^{-2} 
 \left(  \beta^p -  \beta_0 - (\hat{\beta}_{IV, \tau n}(0) - \beta_0)
 \right)  \nonumber  \\
& & + \frac{1}{( n - [\tau n] ) } 
 \left(  \beta^p -  \beta_0 - (\hat{\beta}_{IV, \tau n}(0) - \beta_0)
 \right)'
\left(  
\frac{X_{\tau n}'
P_{Z_{\tau n}} 
 X_{\tau n}}{ [ \tau n ] }  \right)^{-1}  \nonumber 
\\
& & 
\hspace{.5in}
\times  
 X_{n(1-\tau)}'
 P_{Z_{ n(1-\tau)}}
  X_{n(1-\tau)}  \left(  
\frac{X_{\tau n}'
P_{Z_{\tau n}} 
 X_{\tau n}}{ [ \tau n ] }  \right)^{-1} \left(  \beta^p -  \beta_0 - (\hat{\beta}_{IV, \tau n}(0) - \beta_0)
 \right).
\end{eqnarray*}
The first term will converge to zero and the second term converges to the positive value 
$$
 \left(  \beta^p -  \beta_0 \right)'
\left(  \Gamma_0' R_z \Gamma_0  \right) \left(  \beta^p -  \beta_0
 \right).
$$

Now follow the standard approach (Lemma 5.4 Ichimura (1993)) to show that $\sqrt{n}(\hat{\alpha} - \alpha_0) = O_p(1).$
Expand $Q_{n(1-\tau)}(\alpha)$ about $\alpha_0$ and evaluate at $\hat{\alpha}$.
\begin{eqnarray*}
Q_{n(1-\tau)}( \hat{\alpha} ) 
& = & Q_{n(1-\tau)}(\alpha_0) +  \frac{d Q_{n(1-\tau)}(\alpha_0)}{ d \alpha }
(\hat{\alpha} - \alpha_0) + \frac{1}{2} 
 \frac{d^{2} Q_{n(1-\tau)}( \bar{\alpha})}{d \alpha^2}
(\hat{\alpha} - \alpha_0)^{2}
\end{eqnarray*}
where $ 0 \leq \bar{\alpha} \leq \hat{\alpha}$.  Because $\hat{\alpha} = \argmin_{[0, \infty)} Q_{n(1-\tau)}(\alpha)$,  $0  \geq Q_{n(1-\tau)}( \hat{\alpha} ) 
- Q_{n(1-\tau)}(\alpha_0),$ hence 
\begin{eqnarray*}
0 
& \geq &  \frac{d Q_{n(1-\tau)}(\alpha_0)}{ d \alpha }
(\hat{\alpha} - \alpha_0) + \frac{1}{2} 
 \frac{d^{2}Q_{n(1-\tau)}( \bar{\alpha})}{d \alpha^2}
(\hat{\alpha} - \alpha_0)^{2}.
\end{eqnarray*}
Multiply both sides by 
$
\frac{n}{ (  1 + \sqrt{n}| \hat{\alpha} - \alpha_0 | )^2 }.
$

\begin{eqnarray}
\hspace{-.3in} 0 &\geq & \frac{d Q_{n(1-\tau)}(\alpha_0)}{ d \alpha }
(\hat{\alpha} - \alpha_0)  \frac{n}{ (  1 + \sqrt{n}| \hat{\alpha} - \alpha_0 | )^2 }  
+ \frac{1}{2} 
 \frac{d^{2} Q_{n(1-\tau)}( \bar{\alpha})}{d \alpha^2}
(\hat{\alpha} - \alpha_0)^{2} \frac{n}{ (  1 + \sqrt{n}| \hat{\alpha} - \alpha_0 | )^2 }
\nonumber
\\
\hspace{-.3in}  & = & \sqrt{n} \frac{d Q_{n(1-\tau)}(\alpha_0)}{ d \alpha }
\left( \frac{ \sqrt{n} (\hat{\alpha} - \alpha_0)  }{ (  1 + \sqrt{n}| \hat{\alpha} - \alpha_0 | ) } \right) 
 \frac{1}{ (  1 + \sqrt{n}| \hat{\alpha} - \alpha_0 | ) }
+ \frac{1}{2} 
 \frac{d^{2} Q_{n(1-\tau)} ( \bar{\alpha})}{d \alpha^2}
 \left(  \frac{ \sqrt{n} (\hat{\alpha} - \alpha_0)  }{ (  1 + \sqrt{n}| \hat{\alpha} - \alpha_0 | ) } \right)^2
 \label{bob}
\end{eqnarray}

Suppose $ \sqrt{n}|  \hat{\alpha} - \alpha_0 |$ diverged to infinity.  As noted above  $\sqrt{n} \frac{d Q_{n(1-\tau)}(\alpha_0)}{ d \alpha } = O_p(1).$  Also, $\left( \frac{ \sqrt{n} (\hat{\alpha} - \alpha_0)  }{ (  1 + \sqrt{n}| \hat{\alpha} - \alpha_0 | ) } \right) = O_p(1)$. However, $ 
 \frac{1}{ (  1 + \sqrt{n}| \hat{\alpha} - \alpha_0 | ) } = o_p(1)$ and  hence the first term on the LHS of equation (\ref{bob}) goes to zero.  But this means 
\begin{eqnarray*}
o_p(1) &\geq &   
 \frac{1}{2} 
 \frac{d^{2}Q_{n(1-\tau)}( \bar{\alpha})}{d \alpha^2}
 \left(  \frac{ \sqrt{n} (\hat{\alpha} - \alpha_0)}{ (  1 + \sqrt{n}| \hat{\alpha} - \alpha_0 | ) } \right)^2.
\end{eqnarray*}

But $ \frac{d^{2}Q_{n(1-\tau)}( \bar{\alpha})}{d \alpha^2} $ limits to $ \frac{d^{2} Q_{n(1-\tau)}( \alpha_0)}{d \alpha^2} $, a  positive value, and the RHS can satisfy this only if  
$$
\frac{ \sqrt{n} (\hat{\alpha} - \alpha_0)}{ (  1 + \sqrt{n}| \hat{\alpha} - \alpha_0 | ) } = o_p(1).
$$
This occurs only if $\sqrt{n} | \hat{\alpha} - \alpha_0 | = o_p(1)$ which is a contradiction of the assumption that $\sqrt{n}|  \hat{\alpha} - \alpha_0  |$ diverges.   Hence $\sqrt{n} ( \hat{\alpha} - \alpha_0 ) = O_p(1)$. 
\noindent
  \QEDA

\section{Proof of Theorem 1}

This is a direct application of {\bf Theorem 1} from  \citeA{andrews2002generalized}. {\bf Assumptions GMM1$^*$ - GMM5$^*$} in \citeA{andrews2002generalized} are satisfied for the linear model by {\bf Assumptions 1 - 3}.  To  show how the assumptions in  \citeA{andrews2002generalized}  are satisfies, we first use {\bf Assumtions 1 - 3} to demonstrate three useful results for the system of equations (\ref{moment_conditions}).
 The useful results are: $ E[h_i(\theta_0) ] = 0 $, $\sqrt{n} H_n(\theta_0)$ satisfies a central limit theorem and $ \left( \lim_{n \rightarrow \infty} \frac{\partial H_n(\theta_0)}{\partial \theta'} \right)^{-1} $ exists, which requires showing that LLN leads to a matrix which is invertible.
 In the statement of the Theorem, the limiting random variable, $Z$, is composed of two terms:  $\sqrt{n} H_n(\theta_0)$ and $ \left(  - E \left[ \frac{\partial h_{i} (\theta_0)}{\partial \theta'} \right] \right)^{-1}$.
  
  Evaluate the moment condition, equations (\ref{moment_conditions}), at $\theta_0$, to show that $E[h_i(\theta_0)] = 0$ and that $\sqrt{n} H_n(\theta_0)$ satisfies a central limit theorem.

$$
H_n(\theta_0) = \frac{1}{n} \sum_{i=1}^{n}  \left[ 
\begin{array}{c}
{\bf 1}_{\tau n}(i) {\rm vech}(R_z - z_i z_i') \\
{\bf 1}_{\tau n}(i) {\rm vec}( R_z\Gamma_0  - z_i x_i') \\
{\bf 1}_{\tau n}(i)  \left( \Gamma_0' R_z R_z^{-1} z_i ( y_i  - x_i' \beta_{0})   \right) \\
(1 - {\bf 1}_{\tau n}(i) ) (y_i - x_i '\beta_{0}) z_i 'R_{z}^{-1}  R_z\Gamma_0    \left( \Gamma_0' R_z R_z^{-1}  R_z\Gamma_0     \right)^{-1} (\beta^p - \beta_{0})
\\
(1 - {\bf 1}_{\tau n}(i) ) {\rm vech}(R_z - z_i z_i') \\
(1 - {\bf 1}_{\tau n}(i) )  {\rm vec}( R_z\Gamma_0   - z_i x_i') 
\end{array}
\right]
$$
$$
 = \frac{1}{n} \sum_{i=1}^{n}  \left[ 
\begin{array}{c}
{\bf 1}_{\tau n}(i) {\rm vech}(R_z - z_i z_i') \\
{\bf 1}_{\tau n}(i) {\rm vec} \left( R_z\Gamma_0 -   z_i u_i'- z_i z_i'\Gamma_0  \right) 
\\
{\bf 1}_{\tau n}(i)  \left( \Gamma_0' z_i \epsilon_i   \right) \\
(1 - {\bf 1}_{\tau n}(i) ) \epsilon_i z_i' \Gamma_0    \left( \Gamma_0' R_z \Gamma_0     \right)^{-1} (\beta^p - \beta_{0})
\\
(1 - {\bf 1}_{\tau n}(i) ) {\rm vech}(R_z - z_i z_i') \\
(1 - {\bf 1}_{\tau n}(i) )  {\rm vec}(  R_z \Gamma_0 -  u_iz_i'- z_i z_i' \Gamma_0  ) 
\end{array}
\right]
$$
Each element of  $h_i(\theta_0)$  has expectation zero and bounded covariance, hence the iid assumption implies the central limit theorem
\begin{eqnarray}
& & 
\sqrt{n} H_n(\theta_0) \sim^A  \nonumber
\\
& & N\left(0, \left[ \begin{array}{c c} \tau I_{ \left\{ \frac{m(m+1)}{2} + km   + k \right\}} & 0 
\\ 0 & (1-\tau) I_{ \left\{ 1 + \frac{m(m+1)}{2} + km \right\} }   \end{array} \right]  \left[\begin{array}{c c c c c c c} \chi & \xi & 0 & 0 & 0 & 0 
\\
\xi' & \zeta & \Psi & 0 & 0 & 0 
\\
0 & \Psi' & \Xi & 0 & 0 & 0 
\\
0 & 0 & 0 & \Upsilon & 0 & \Pi 
\\
0 & 0 &  0 & 0 & \chi & \xi 
\\
0 & 0 &  0 & \Pi' & \xi' & \zeta  
\end{array} \right] \right)  
\nonumber % \label{CLT}
\end{eqnarray}
where
$$
  \chi = E \left[  {\rm vech}(R_z - z_i z_i')
   {\rm vech}(R_z - z_i z_i')'
 \right],
$$
$$
\xi = E\left[   {\rm vech}(R_z - z_i z_i')   {\rm vec}(R_z \Gamma_0 - 
z_i z_i' \Gamma_0 )'    \right], 
$$
$$
\zeta = E \left[ \ {\rm vec}( R_z \Gamma_0 -z_i x_i') {\rm vec}( R_z \Gamma_0 - z_i x_i' )' \right],
$$ 

$$
 \Psi= E\left[ 
 {\rm vec}(  z_i u_i' )
   \left( \epsilon_i z_i' \Gamma_0     \right) 
\right],
$$
$$
\Xi =  ( \Gamma_0'  R_z \Gamma_0  ) \sigma_\epsilon^2,
$$ 
$$
 \Upsilon= \sigma_\varepsilon^2 
 (\beta^p - \beta_{0})' ( \Gamma_0' R_z \Gamma_0 )^{-1} (\beta^p - \beta), \mbox{ and}
$$ 
$$
\Pi\ = E \left[    \varepsilon_i z_i '\Gamma_{0} ( \Gamma_0' R_z \Gamma_0 )^{-1} (\beta^p - \beta_{0})
  {\rm vec}( -  u_iz_i' )'
\right].
$$ 

The expectation of the first derivative of the moment conditions evaluated at $\theta_0$ is
\begin{eqnarray*}
& & E \left[ \frac{\partial h_i(\theta_{0})}{\partial \theta'} \right] 
 = 
%\\
%
%& & 
  \left[ \begin{array}{c c} \tau I_{ \left\{ \frac{m(m+1)}{2} + km     +k \right\}} & 0 
\\ 0 & (1-\tau) I_{ \left\{ 1 + \frac{m(m+1)}{2} + km \right\} }   \end{array} \right]  
\left[ 
\begin{array}{c c c}
I_{ \left\{ \frac{m(m+1)}{2} + km \right\}} & 0 & 0 
\\
0 & D & 0 
\\
0 & 0 & I_{ \left\{ \frac{m(m+1)}{2} + km \right\}} 
\end{array}
\right]
\end{eqnarray*}

% { 
% %\tiny 
% \scriptsize 
% %\footnotesize
% \begin{eqnarray*}
% \hspace{-.9in}  \left[\begin{array}{c c  c c c c c}
% I_{\left\{  \frac{m(m+1)}{2} \right\}} & 0 & 0 & 0 & 0 & 0 & 0  
% \\
% 0 & I_{km} & 0 & 0 & 0 & 0 & 0  
% \\
% \hline
% \left\{ {\rm  vech}( \Gamma_0' \frac{\partial R_\tau}{\partial r_j} \Gamma_0 )  \right\}_{j=1,\ldots,\frac{m(m+1)}{2} } & - \left\{   {\rm  vech} \left( \frac{\partial S_\tau'}{\partial s_j} \Gamma_0 + \Gamma_0' \frac{\partial S_\tau}{\partial s_j} \right)  \right\}_{j=1,\ldots, km} & I_{\left\{  \frac{k(k+1)}{2} \right\}} & 0  & -{\rm  vech}( I_k) & 0 & 0 
% \\
%  \left\{ \Omega_{0}^{-1}  \Gamma_0'  \frac{\partial R_\tau}{\partial r_j} \Gamma_0  \beta_0 \right\}_{j=1,\ldots,\frac{m(m+1)}{2} }   & - \left\{ \Omega_{0}^{-1} \frac{\partial S_\tau'}{\partial s_j} \Gamma_0  \beta_0 \right\}_{j=1, \ldots, km} &   \left\{ \Omega_{0}^{-1} \frac{\partial \Omega}{\partial \omega_j} \beta_0 \right\}_{j=1,\ldots, \frac{k(k+1)}{2} } & I_k  & -\Omega_{0}^{-1} \beta^p & 0 & 0 
% \\
% 0 & 0  & 0 & - (\beta^p - \beta_{0})'_{} & 0 & 0 & 0  
% \\
% 0 & 0 & 0 & 0 & 0 & I_{\left\{  \frac{m(m+1)}{2} \right\}} & 0  
% \\
% 0 & 0 & 0 & 0 & 0 & 0 & I_{km}  
% \end{array} \right] 
% \end{eqnarray*} 
%  
% } % end the tiny font  
%
where
$$
D  = \left[ \begin{array}{c c}  
-S_0 'R_z^{-1} S_0 & (\beta_0 - \beta^p)  
\\
(\beta_0 - \beta^p)'   & 0  
    \end{array} \right].
$$

The inverse is well defined by {\bf Assumption 3} and given by 
$$
\left( 
E \left[ \frac{\partial h_i(\theta_{0})}{\partial \theta'} \right] 
\right)^{-1}
=
\left[ 
\begin{array}{c | c c}
I_{ \left\{ \frac{m(m+1)}{2} + km \right\}} & 0 & 0 
\\
\hline
0 & D^{-1} & 0 
\\
0 & 0 & I_{ \left\{ \frac{m(m+1)}{2} + km \right\}} 
\end{array}
\right]
$$
where
$$
D^{-1}  = \frac{1}{ \tilde{\delta} } \left[ \begin{array}{c c}  
-\tilde{\delta} \left(S_0 'R_z^{-1} S_0\right)^{-1} +
 \left( S_0 'R_z^{-1} S_0 \right)^{-1} (\beta_0 - \beta^p)
 (\beta_0 - \beta^p)'  \left( S_0 'R_z^{-1} S_0 \right)^{-1}    &  \left( S_0 'R_z^{-1} S_0 \right)^{-1} (\beta_0 - \beta^p)     
\\
   (\beta_0 - \beta^p)'  \left( S_0 'R_z^{-1} S_0 \right)^{-1}  &   1
    \end{array} \right]
$$
and $\tilde{\delta} = (\beta_0 - \beta^p)' \left( S_0 'R_z^{-1} S_0 \right)^{-1} (\beta_0 - \beta^p). $   
Hence $ \left(  - E \left[ \frac{\partial h_{i} (\theta_0)}{\partial \theta'} \right] \right)^{-1}$ is well defined. 
Now verify {\bf Assumptions    GMM1$^*$  -  GMM5$^*$  }in Anderws (2002).

\noindent
{\bf Assumption GMM1$^*$: }This parameter space is bounded. Because $ z_i $ has finite fourth moments and $\left[ 
 \begin{array}{c c}
 \varepsilon_i 
 &
 u_i' 
 \end{array}
 \right]'$ has a finite second moment there exists a dominating function with a finite expectation.  This implies that $H_n(\theta)' H_n(\theta) $ will uniformly converge to its limiting function, $E [H_n(\theta)'] E [H_n(\theta)]$. Identification follows from $E [H_n(\theta_0)] = 0$ and the invertibility of $ M_0. $

\bigskip
\noindent
{\bf Assumption GMM2$^*$: }The data are iid. The GMM structure is presented above. The expectation of the first derivative of the moment conditions is evaluated at $\theta_0$ and inverted, hence demonstrating it is full rank.   $E [H_n(\theta_0)] = 0$ is demonstrated above.  The system is just identified,  so an identity weighting matrix is used.

\bigskip
\noindent
{\bf Assumption GMM3$^*$: }The CLT applies because the data are iid and  $ z_i $ has finite fourth moments, $\left[ 
 \begin{array}{c c}
 \varepsilon_i 
 &
 u_i' 
 \end{array}
 \right]'$ has a finite second moment and the $z_i$ and $\left[ 
 \begin{array}{c c}
 \varepsilon_j 
 &
 u_j' 
 \end{array}
 \right]'$ are independent for all $i$ and $j$.

\bigskip
\noindent
{\bf Assumption GMM4$^*$: }Because the eigenvalues of $R_z$ are bounded above zero and below infinity each element of $R_z$ and $R_z^{-1}$ is bounded above.  Hence all the parameters in $\Theta$ are bounded and equation (27) of \citeA{andrews2002generalized} is satisfied with $c = \max(B_1, B_2, B_3, B_4) $.  

\bigskip
\noindent
{\bf Assumption GMM5$^*$: } The cone for this problem is 
$
\Lambda = \left\{ \lambda \in R^{ m(m+1) + 2 mk + \frac{k(k+1)}{2} + k + 1  }: \lambda_{ \frac{m(m+1)}{2} + mk + \frac{k(k+1)}{2} + k +1} \geq 0 \right\}
$
which is convex.
\noindent
  \QEDA

\end{document}